\documentclass[prd,preprint,tightenlines,floatfix,showpacs,preprintnumbers
,nofootinbib,superscriptaddress,eqsecnum]{revtex4}

 \usepackage[dvips,final]{graphicx}
  \usepackage{amssymb}
   \usepackage{amsmath}
    \usepackage{amsfonts}
     \usepackage{epsfig}
      \usepackage{bm}
      \usepackage{xcolor}

\newcommand{\bfB}{\mbox{\boldmath $B$}}

\newcommand{\bPhi}{\mbox{\boldmath $\Phi$}}
\newcommand{\bkappa}{\mbox{\boldmath $\kappa$}}
\newcommand{\bl}{\mbox{\boldmath $l$}}

\newcommand{\bq}{\mbox{\boldmath $q$}}
\newcommand{\bn}{\mbox{\boldmath $n$}}

\newcommand{\ba}{\mbox{\boldmath $a$}}

\newcommand{\bk}{\mbox{\boldmath $k$}}
\newcommand{\bfb}{\mbox{\boldmath $b$}}

\newcommand{\Jot}{{\cal{J}}}

\def\cpc#1#2#3  {{Computer\ Phys.\ Comm.\ }  {\bf#1}, #2 (#3)}
\def\err#1#2#3  {{\it Erratum }              {\bf#1}, #2 (#3)}
\def\epjc#1#2#3 {{Eur. Phys. J. C }          {\bf#1}, #2 (#3)}
\def\dum#1#2#3  {{~}                         {\bf#1}, #2 (#3)}
\def\ib#1#2#3   {{\it ibid. }                {\bf#1}, #2 (#3)}
\def\jcp#1#2#3  {{J.\ Comput.\ Phys.\ }      {\bf#1}, #2 (#3)}
\def\jetpl#1#2#3 {{\rm JETP Lett.}           {\bf#1}, #2 (#3)}
\def\jhep#1#2#3 {{JHEP }                     {\bf#1}, #2 (#3)}
\def\ijmp#1#2#3 {{Int.\ J.\ Mod.\ Phys.\ }   {\bf#1}, #2 (#3)}
\def\jpg#1#2#3  {{J.\ Phys.\ G }             {\bf#1}, #2 (#3)}
\def\mpl#1#2#3  {{Mod.\ Phys.\ Lett.\ }      {\bf#1}, #2 (#3)}
\def\ncim#1#2#3 {{Nuovo Cimento }            {\bf#1}, #2 (#3)}
\def\np#1#2#3   {{Nucl.\ Phys.\ }            {\bf#1}, #2 (#3)}
\def\pan#1#2#3  {{Phys.\ At.\ Nuclei }       {\bf#1}, #2 (#3)}
\def\plb#1#2#3  {{Phys.\ Lett.\ B }          {\bf#1}, #2 (#3)}
\def\prep#1#2#3 {{Phys.\ Rep.\ }             {\bf#1}, #2 (#3)}
\def\prd#1#2#3  {{Phys.\ Rev.\ D }           {\bf#1}, #2 (#3)}
\def\prd#1#2#3  {{Phys.\ Rev.\ D }           {\bf#1}, #2 (#3)}
\def\prc#1#2#3  {{Phys.\ Rev.\ C }           {\bf#1}, #2 (#3)}
\def\ptp#1#2#3  {{Prog.\ Theor.\ Phys.\ }    {\bf#1}, #2 (#3)}
\def\ps#1#2#3   {{Phpsica Scripta }          {\bf#1}, #2 (#3)}
\def\rmp#1#2#3  {{Rev.\ Mod.\ Phys.\ }       {\bf#1}, #2 (#3)}
\def\rpp#1#2#3  {{Rep.\ Prog.\ Phys.\ }      {\bf#1}, #2 (#3)}
\def\sa#1#2#3   {{Sci. Acta}                 {\bf#1}, #2 (#3)}
\def\sjnp#1#2#3 {{Sov.\ J.\ Nucl.\ Phys.\ }  {\bf#1}, #2 (#3)}
\def\spj#1#2#3  {{Sov.\ Phys.\ JETP }        {\bf#1}, #2 (#3)}
\def\spjl#1#2#3 {{Sov.\ JETP Lett.\ }        {\bf#1}, #2 (#3)}
\def\spu#1#2#3  {{Sov.\ Phys.-Usp.\ }        {\bf#1}, #2 (#3)}
\def\yaf#1#2#3  {{Yad.\ Fiz.\ }              {\bf#1}, #2 (#3)}
\def\zp#1#2#3   {{Zeit.\ Phys.\ }            {\bf#1}, #2 (#3)}
\def\zpc#1#2#3  {{Z.\ Phys.\ C }             {\bf#1}, #2 (#3)}

\def\lsim{\mathrel{\rlap{\lower4pt\hbox{\hskip1pt$\sim$}}
    \raise1pt\hbox{$<$}}}         
\def\gsim{\mathrel{\rlap{\lower4pt\hbox{\hskip1pt$\sim$}}
    \raise1pt\hbox{$>$}}}         
\begin{document}

\title{Two-gluon exchange contribution to elastic
$\gamma \gamma \to \gamma \gamma$ scattering \\ 
and production of two-photons in \\ 
ultraperipheral ultrarelativistic heavy ion \\
and proton-proton collisions}

\author{Mariola K{\l}usek-Gawenda}
\email{mariola.klusek@ifj.edu.pl}
\affiliation{Institute of Nuclear
Physics PAN, PL-31-342 Cracow, Poland}

\author{Wolfgang Sch\"afer}
\email{wolfgang.schafer@ifj.edu.pl} 
\affiliation{Institute of Nuclear
Physics PAN, PL-31-342 Cracow, Poland}

\author{Antoni Szczurek}
\email{antoni.szczurek@ifj.edu.pl} 
\affiliation{Institute of Nuclear
Physics PAN, PL-31-342 Cracow,
Poland}
\affiliation{
University of Rzesz\'ow, PL-35-959 Rzesz\'ow, Poland}

\date{\today}

\begin{abstract}
We discuss the two-gluon exchange contribution (formally three-loops) 
to elastic photon-photon scattering in the high-energy approximation.
The elastic $\gamma\gamma \to \gamma \gamma$ amplitude is given in the impact-factor representation
for all helicity configurations and finite quark masses.
We discuss the importance of including the charm quark, 
which contribution, due to interference, can enhance the cross section
considerably. 
We investigate the contribution to the $\gamma \gamma \to \gamma \gamma$ amplitude 
from the soft region, by studying its dependence 
on nonperturbative gluon mass.
Helicity-flip contributions are shown to be much smaller than
helicity-conserving ones. We identify region(s) of phase space where 
the two-gluon exchange contribution becomes 
important ingredient compared to box and nonperturbative 
VDM-Regge mechanisms considered in the literature.
Consequences for the $A A \to A A \gamma \gamma$ reaction are
discussed. Several differential distributions are shown.
A feasibility study to observe the effect of two-gluon exchange is presented.
We perform a similar analysis for the $p p \to p p \gamma \gamma$
reaction. Only by imposing severe cuts on $M_{\gamma \gamma}$ and
a narrow window on photon transverse momenta the two gluon contribution
becomes comparable to the box contribution but the corresponding cross
section is rather small.
\end{abstract}


\maketitle

\section{Introduction}

Recently a possibility of measuring
elastic photon-photon scattering was discussed for the first time
\cite{Enterria,KLS2016}.
Especially the recent calculation \cite{KLS2016}, which found a 
substantially larger cross section than earlier estimates, has rekindled the
interest of LHC experiments.
In this previous study of two of us \cite{KLS2016}, 
we have considered scattering via a fermion or $W^+ W^-$ loop 
(the so-called box mechanisms) as well as a nonperturbative mechanism 
of fluctuation of both photons into vector mesons 
and their subsequent interaction.

The second mechanism, which involves the 
Reggeon and Pomeron exchanges between vector mesons, 
leads to a rising elastic $\gamma \gamma$ cross section (see also 
\cite{Klusek:2009yi} for the related $\rho^0 \rho^0$ final state).
Fermion boxes, due to the lower spin exchanged in the crossed channels,
drop as a function of energy. 
The $W^+ W^-$-box, which gives a flat energy dependence becomes 
relevant only at large invariant masses of the diphoton system, 
$M_{\gamma \gamma} \gsim 2 m_W$.

The hadronic Pomeron exchange contribution may dominate over the 
box mechanisms only at high subsystem
energies, when the large contribution from the fermion boxes has died out, 
which means large rapidity distances between photons 
in heavy ion collisions. 

Here we consider another mechansism which gives rise to a flat 
cross section at high $\gamma \gamma$ center of mass energy:
the exchange of two gauge bosons between fermion-loops.
In practice we restrict ourselves to the dominant two-gluon exchange
contribution. 

\begin{figure}[!h]
\includegraphics[scale=0.45]{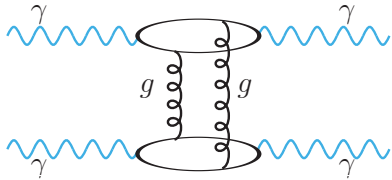}
\caption{Elementary $\gamma\gamma \to \gamma\gamma$ processes 
via two-gluon exchange discussed in extenso in the present
paper. 
}
\label{fig:two-gluon_exchange}
\end{figure}

Formally the two-gluon exchange mechanism shown in Fig.~\ref{fig:two-gluon_exchange}
is a three-loop mechanism.
Its contribution to the elastic scattering of photons at high energies
has been first considered in the pioneering work \cite{Ginzburg:1985tp}.
Indeed in the limit where the Mandelstam variables of the 
$\gamma \gamma \to \gamma \gamma$ process satisfy $\hat s \gg -\hat t, - \hat u$,
major simplifications occur and the three-loop process becomes tractable.
This corresponds to a near-forward, small-angle, scattering of photons.
  
In our treatment, we go beyond the early work \cite{Ginzburg:1985tp} 
by including finite fermion masses, as well as the full momentum
structure in the loops, and we consider all helicity amplitudes. 

The applicability of perturbative QCD (pQCD) 
requires a dominance of short distances, which should be 
ensured by a hard scale. As we deal with real photons, we are
required to ask for a large momentum transfer, say 
$-\hat t, \ -\hat u  \gg 1 \, \rm{GeV}^2$. 
 
The renewed interest to study $\gamma \gamma \to \gamma \gamma$ 
in heavy ion collsions
makes the analysis of off-forward amplitude rather topical.

\begin{figure}[!h]
\includegraphics[scale=0.4]{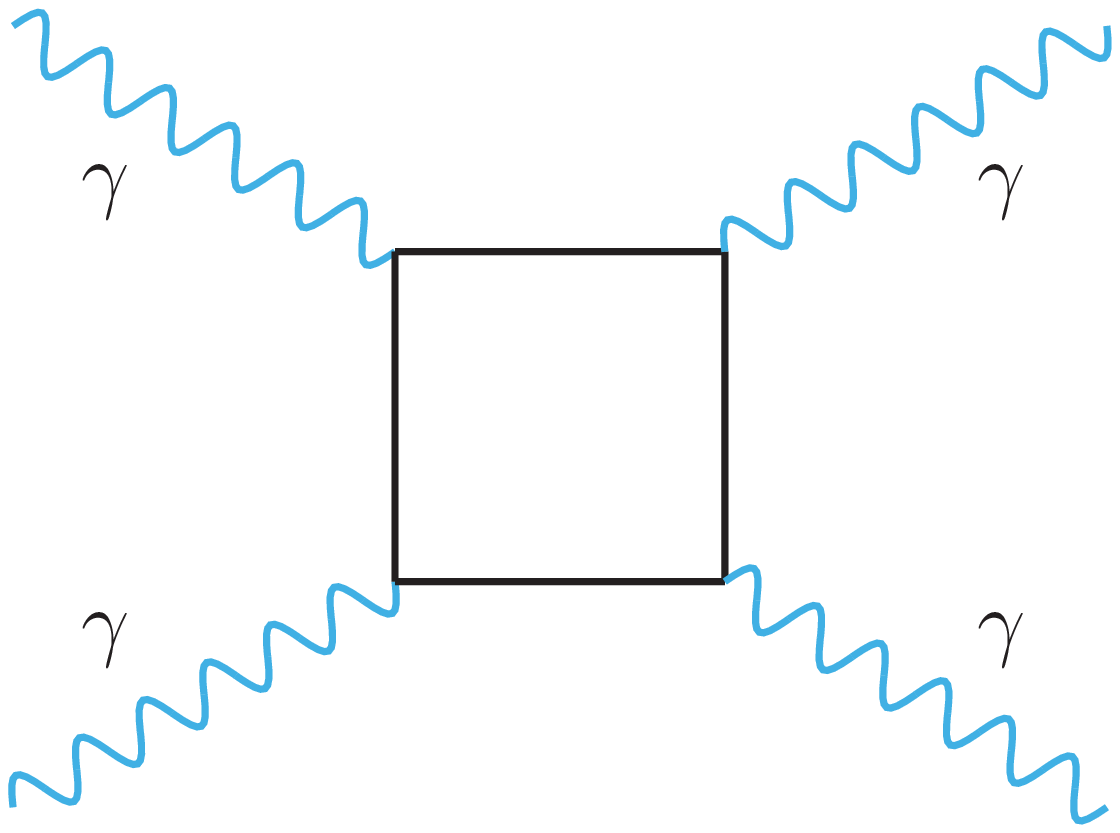}
\includegraphics[scale=0.45]{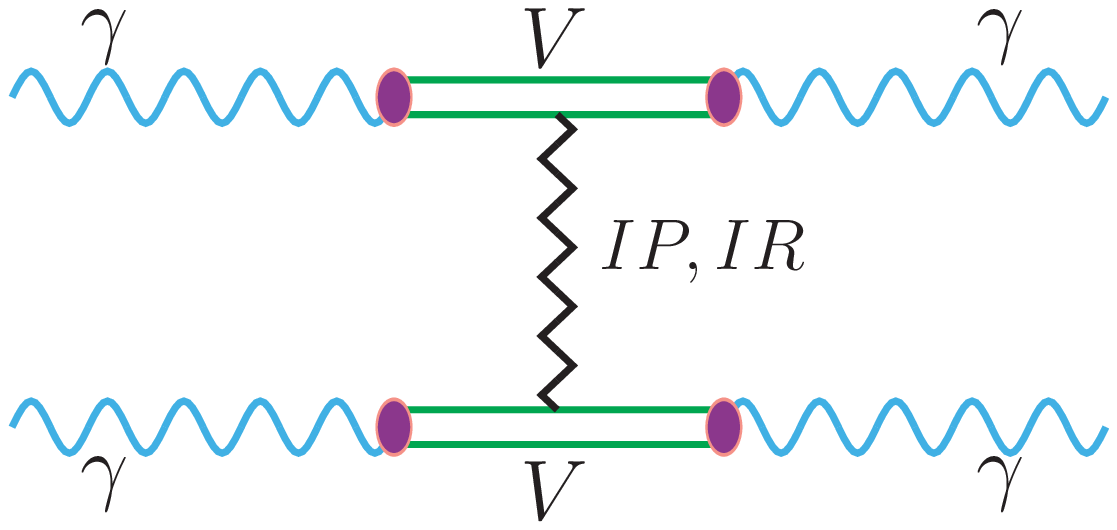}
\caption{Other elementary $\gamma\gamma \to \gamma\gamma$ processes. 
The left panel represents the box mechanism and the right panel 
is for VDM-Regge mechanism considered recently in Ref.\cite{KLS2016}.}
\label{fig:other_mechanism}
\end{figure}

For reference we shall consider also the box mechansism (see left panel
of Fig.~\ref{fig:other_mechanism}) and
the VDM-Regge mechanism (see right panel of Fig.~\ref{fig:other_mechanism}.) 
where photons fluctuate into virtual vector mesons
(three different light vector mesons are included). 
In this case the interaction ''between photons'' 
happens when both photons are in their hadronic states.
The latter mechanism has very similar kinematics
as the two-gluon mechanism discussed in the present paper in detail,
but is concentrated at very small momentum transfers.

\section{Theoretical approach}

\subsubsection{$\gamma \gamma \to \gamma \gamma$ elastic scattering}

\begin{figure}[!h]
\includegraphics[scale=0.5]{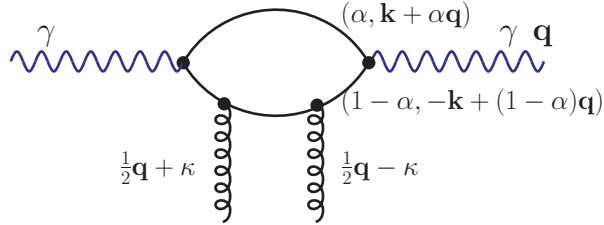}
\caption{
Kinematical variables used in calculating elementary $\gamma\gamma \to \gamma\gamma$ processes 
via two-gluon exchange. 
}
\label{fig:two-gluon_exchange_notation}
\end{figure}

The altogether 16 diagrams of the type shown in 
Fig.~\ref{fig:two-gluon_exchange} and Fig.~\ref{fig:two-gluon_exchange_notation} result in the amplitude, which can be
cast into the impact-factor representation \cite{Cheng:1970ef}:
\begin{eqnarray}
{\cal M}(\gamma_{\lambda_1} \gamma_{\lambda_2} 
\to \gamma_{\lambda_3} \gamma_{\lambda_4}; \hat s, \hat t) = i \hat s \sum_{f,f'}^{n_f} 
\int d^2 \bkappa \, 
{ \Jot^{(f)}( \gamma_{\lambda_1} \to \gamma_{\lambda_3}; \bkappa,\bq) 
\,  \Jot^{(f')}( \gamma_{\lambda_2} \to \gamma_{\lambda_4}; - \bkappa, - \bq)
\over
[(\bkappa + \bq/2)^2 + m_g^2][(\bkappa-\bq/2)^2 + m_g^2]
}  \, . \nonumber \\
\label{eq:A_2g}
\end{eqnarray}
Here $\bq$ is the transverse momentum transfer, $\hat t \approx - \bq^2$, and
$m_g$ is a gluon mass parameter, which role will be discussed below. 
We parametrize the loop momentum such that 
gluons carry transverse momenta $\bq/2 \pm \bkappa$ 
(see Fig.~\ref{fig:two-gluon_exchange_notation}). 
Notice, that the amplitude is finite at $m_g \to 0$, because 
the impact factors $\Jot$ vanish for $\bkappa \to \pm \bq/2$.

The amplitude is normalized such, that 
the differential cross section is given by
\begin{eqnarray}
{d \sigma(\gamma \gamma \to \gamma \gamma; \hat s) \over d \hat t} = {1 \over 16 \pi \hat s^2} 
\, {1 \over 4} \sum_{\lambda_i} \Big| 
{\cal M} (\gamma_{\lambda_1} \gamma_{\lambda_2} 
\to \gamma_{\lambda_3} \gamma_{\lambda_4}; \hat s, \hat t) \Big|^2
\, .
\end{eqnarray}
In this case, the explicit form of the impact factor is
\begin{eqnarray}
&&\Jot^{(f)}( \gamma_{\lambda} \to \gamma_{\tau}; \bkappa,\bq) =
 \, \sqrt{N_c^2-1} \,  \,
\,  {e^2_f \alpha_{\rm{em}} \over 2 \pi^2} \, \int_0^1 d\alpha \int {d^2 \bk \over \bk^2 + m_f^2} \alpha_S(\mu^2) \, \nonumber \\
&&\times \Big\{ \delta_{\lambda \tau} 
\Big( m_f^2 \, \Phi_2 + 
\left[ \alpha^2 + (1-\alpha)^2 \right] (\bk\bPhi_1) \Big) 
+ \delta_{\lambda,-\tau} 
2\alpha(1-\alpha) \Big( (\bPhi_1 \bn)(\bk \bn) - [\bPhi_1 , \bn] [\bk,\bn] \Big)  \Big\} \, . \nonumber \\ 
\label{eq:explicit_impact_factor}
\end{eqnarray}
Here, $\bn = \bq/|\bq|$, and $[\ba,\bfb] = a_x b_y - a_y b_x$.
Furthermore, $N_c=3$ is the number of colors, $e_f$ is the charge of the 
quark of flavour $f$.
Quark and antiquark share the large lightcone momentum of the incoming photon in
fractions $\alpha,1-\alpha$ respectively. 
The helicity conserving part is easily obtained, after due change of the final state wave function, 
from the one used in the $\gamma \gamma \to J/\psi J/\psi$
process in \cite{Baranov:2012vu}. Also the helicity-flip piece can be obtained, mutatis mutandis,
from the $\gamma \to V$ impact factors for vector meson final states \cite{Ivanov:2004ax}. 

Above $\bPhi_1, \Phi_2$ are shorthand notations for the momentum structures,
corresponding to the four relevant Feynman diagrams:
\begin{eqnarray}
\Phi_2 = -{1 \over (\bl + \bkappa)^2 + m_f^2}  
-{1 \over (\bl - \bkappa)^2 + m_f^2} 
+{1 \over (\bl + \bq/2)^2 + m_f^2}
+{1 \over (\bl - \bq/2)^2 + m_f^2} \;,
\nonumber \\
\bPhi_1 = -{\bl + \bkappa \over (\bl + \bkappa)^2 + m_f^2}  
-{\bl - \bkappa \over (\bl - \bkappa)^2 + m_f^2} 
+{\bl + \bq/2 \over (\bl + \bq/2)^2 + m_f^2}
+{\bl - \bq/2 \over (\bl - \bq/2)^2 + m_f^2}
\, ,
\label{eq:impact_factors}
\end{eqnarray}
and we have used 
\begin{eqnarray}
\bl = \bk + \Big( \alpha - {1 \over 2} \Big)\bq \, . 
\end{eqnarray}

In the present approach we assume incoming real photons and therefore 
only transverse photon polarizations are taken into account.
This is a sufficiently good approximation for heavy-ion peripheral
collisions where the nucleus charge form factor selects quasi-real
photons.

The running scale of strong coupling constant for the evaluation of the
two-gluon exchange cross section is taken as:
\begin{eqnarray}
\mu^2 &=& \max\{\bkappa^2,\bk^2+m_Q^2, \bq^2 \} \; . \nonumber  \\ 
\label{renormalization_scale} 
\end{eqnarray}
We freeze the running coupling in the infrared at 
a value of $\alpha_S \sim 0.8$.

\subsection{$A A \to A A \gamma \gamma$ reaction}

As in our recent analysis \cite{KLS2016} also here
we follow the impact-parameter equivalent photon approximation \cite{Baur:1990fx}, 
called in the following ``b-space EPA'' for brevity.
In this approximation the cross section can be written as: 
\begin{eqnarray}
\sigma_{A_1 A_2 \to A_1 A_2 \gamma \gamma}(s_{A_1 A_2}) 
&=&
\int d^2 \bfB \,  d^2 \bfb \, {d \omega_1 \over \omega_1} {d\omega_2 \over \omega_2} 
\, \sigma_{\gamma \gamma \to \gamma \gamma} (\hat s) 
N\Big(\omega_1, \bfB + {\bfb \over 2} \Big)
N\Big(\omega_2, \bfB - {\bfb \over 2} \Big) 
S_{abs}^2(\bfb)
\;. \nonumber \\
\label{eq:EPA_sigma_final_5int}
\end{eqnarray}
Here $\hat s = M_{\gamma \gamma}^2 = 4 \omega_1 \omega_2$, and
$N(\omega_i,\bfb_i)$ are the photon fluxes in one or second nucleus.
Nuclear charge form factors are the main ingredients of the photon flux.
In our calculations we use a realistic form factor
which is a Fourier transform of a charge distribution in nuclei. More details
about choice of the form factor and on derivation of Eq. (\ref{eq:EPA_sigma_final_5int})
one can find in Ref.\cite{KS_muons}.

The gap survival factor, describing probability that the nucleus
would not undergo break up, to a good approximation, can be written as
(\cite{Baur:1990fx,Klusek:2009yi,KLS2016})
\begin{equation}
S^2_{abs} (\bfb) = \theta \left( |\bfb| - 2 R_A ) \right) \;.
\end{equation}
Only some differential distributions can be calculated from
formula (\ref{eq:EPA_sigma_final_5int}).
To make real comparison to future experimental data or made
predictions for real experiments an inclusion of kinematical variables
of individual photons is necessary. The corresponding 
details have been explained in Ref.\cite{KLS2016} and will be not
repeated here.

\subsection{$p p \to p p \gamma \gamma$ reaction}

In this paper we shall consider also the mechanism of elastic 
photon-photon scattering in $p p \to p p \gamma \gamma$ reaction.
Here, in our exploratory study, we neglect the gap survival factor.
Then the cross section of $\gamma\gamma$ production (via $\gamma \gamma$
fusion) in $pp$ collisions takes the simple parton model form
\begin{equation}
\frac{d \sigma}{d y_1 d y_2 d^2 p_t}
= \frac{1}{16 \pi^2 {\hat s}^2} \, 
x_1 \gamma^{(\rm{el})}(x_1)
x_2 \gamma^{(\rm{el})}(x_2)
\overline{|{\cal M}_{\gamma\gamma \to \gamma\gamma} |^2} \,.
\label{inclusive_gamgam}
\end{equation}
Here $y_1,y_2$ are the rapidities of final state photons, $p_t$ is the
photon transverse momentum, and
\begin{eqnarray}
 x_{1,2} = {p_t \over \sqrt{s}} ( \exp(\pm y_1) + \exp(\pm y_2) ) \, .
\end{eqnarray}
In the numerical calculations for the elastic fluxes we shall use a practical
parametrization of Ref. \cite{DZ1989}. 

However, we should remember that for the proton-proton reaction 
another diffractive QCD mechanism takes place, the Pomeron-Pomeron fusion
in which each Pomeron is treated as QCD ladder.
As shown e.g. in \cite{LPS2014} only at large invariant masses the
$\gamma \gamma \to \gamma \gamma$ mechanism with intermediate boxes could
win with the diffractive mechanism.
Here we wish to analyze how the situation changes when the two-gluon
exchange mechanism is taken into account.
\section{Numerical results}

In our calculations here we take:
$m_u, m_d$ = 0.15 GeV, $m_s$ = 0.3 GeV and $m_c$ = 1.5 GeV.
These are effective masses often used in dipole model calculations.
These are parameters which allow to describe the cross section
$\sigma_{\gamma p}$ even for quasi-real photons \cite{Nikolaev}.
As far as $m_g$ regularization parameter is considered we take
two values: $m_g =$ 0.0, 0.75 GeV. The first value is as for usual
gluon exchange while the second one is suggested by
lattice QCD \cite{lattice_QCD} and the color-dipole analysis of
high energy scattering, see e.g. \cite{Fiore:2014hga} 
and references therein.

\subsection{$\gamma \gamma \to \gamma \gamma$ scattering}

In elastic $\gamma \gamma \to \gamma \gamma$ scattering all quark-loops 
contribute coherently in both impact factors 
(add up algebraically in the impact factors,
which is squared then in the cross section). This means that adding only 
one flavour more changes the result considerably. This is particularly
true for the charm quarks/antiquarks. 
The coherent effect is potentially large, much
larger than for the total cross section where simple algebraic adding
in each impact factor takes place. 
%
\begin{figure}[!ht]
\includegraphics[width=0.35\textwidth]{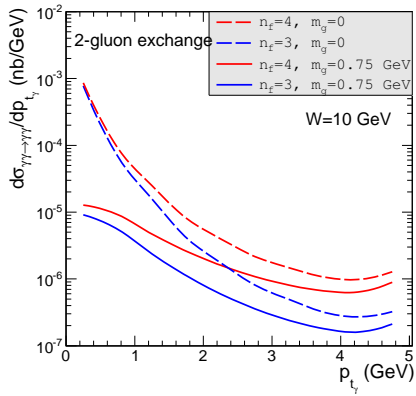}
\caption{Dependence on number of flavours included in the calculation
of transverse momentum distribution of one of outgoing photons for the 
$\gamma \gamma \to \gamma \gamma$ elastic scattering.
The result with three flavours is shown by the blue line, while
that for four flavours by the red line.
}
\label{fig:number_of_flavours}
\end{figure}

In our calculation described above (Eq. (\ref{eq:A_2g})) 
$n_f$ is left as a free parameter.
Here we wish to discuss how our results depend on the number of flavours, $n_f$, 
included in the calculation. An example
for $W =$ 10 GeV is shown in Fig.~\ref{fig:number_of_flavours}.
The results for three flavours are denoted by the blue lines
and the results for four flavours by the red lines. In addition, we show 
distributions for vanishing and finite $m_g$.
The figure shows that inclusion of four flavours is necessary.
In general, the effect increases at larger transverse momenta.

The gluon mass has a large effect in a broad range of $p_t$, and very large
 $p_t$ are necessary for convergence to the massless gluon pQCD limit
($m_g=0$ - dashed lines, $m_g=750$ MeV - solid lines).
A similar observation was made in Ref.~\cite{SNS2022} for pion-pion
elastic scattering.

\begin{figure}[!ht]
\includegraphics[width=0.40\textwidth]{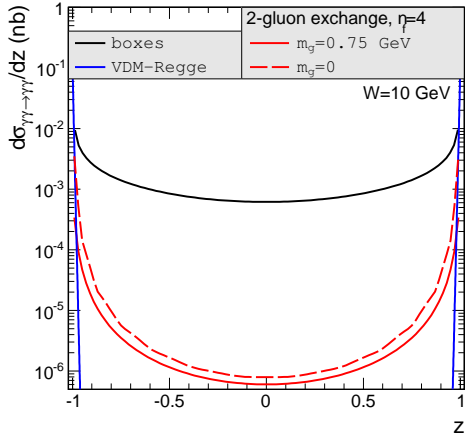}
\label{fig:dsig_dz_10GeV_3processes}
\caption{Competition of the three considered processes 
for $W =$ 10 GeV.
}
\label{fig:dsig_dz_10GeV_3processes}
\end{figure}

Having fixed number of flavours we can focus on the role
of the new mechanism.
How important is the two-gluon contribution compared to
the box and VDM-Regge contributions considered in \cite{KLS2016}
is illustrated in Fig.~\ref{fig:dsig_dz_10GeV_3processes} for relatively 
low energy. Here the cross section differential in $z = \cos \theta$, where
$\theta$ is the scattering angle in the $\gamma \gamma$ cms, is shown.
The contribution of the VDM-Regge is concentrated at
$z \approx \pm$ 1. In contrast, the box contribution extends over a broad
range of $z$. The two-gluon exchange contribution occupies intermediate
regions of $z$.
We need to add though, that the approximations made in the calculation
of the two-gluon exchange are justified in a small angle region only.
At small $z$ the error can easily be 100\%.

\begin{figure}[!ht]
\includegraphics[width=0.40\textwidth]{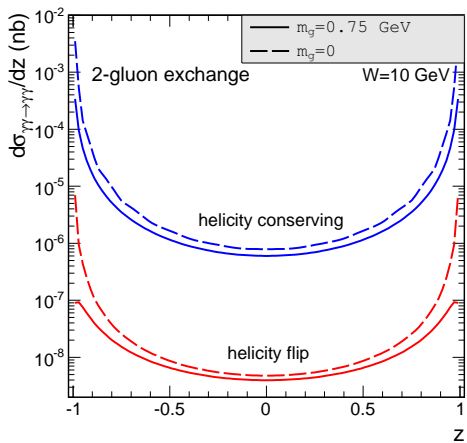}
\caption{Comparison of helicity-conserving and helicity flip contributions
for $W =$ 10 GeV.
}
\label{fig:dsig_dz_10GeV_sf}
\end{figure}

In Fig.~\ref{fig:dsig_dz_10GeV_sf} we show a difference
between the case when $s$-channel-helicity is conserved (upper lines)
and for helicity-flip piece (see Eq.~(\ref{eq:explicit_impact_factor})).
The calculations show that the helicity-flip contributions are about three orders of magnitude smaller
than the $s$-channel helicity conserving pieces.
Note, that due to the zero mass of the photons, a helicity flip involves
two units, the $\Delta \lambda = \pm 1$ 
processes which have some relevance in the 
vector meson production \cite{Ivanov:2004ax} are absent here.

\begin{figure}[!ht]
\includegraphics[width=0.3\textwidth]{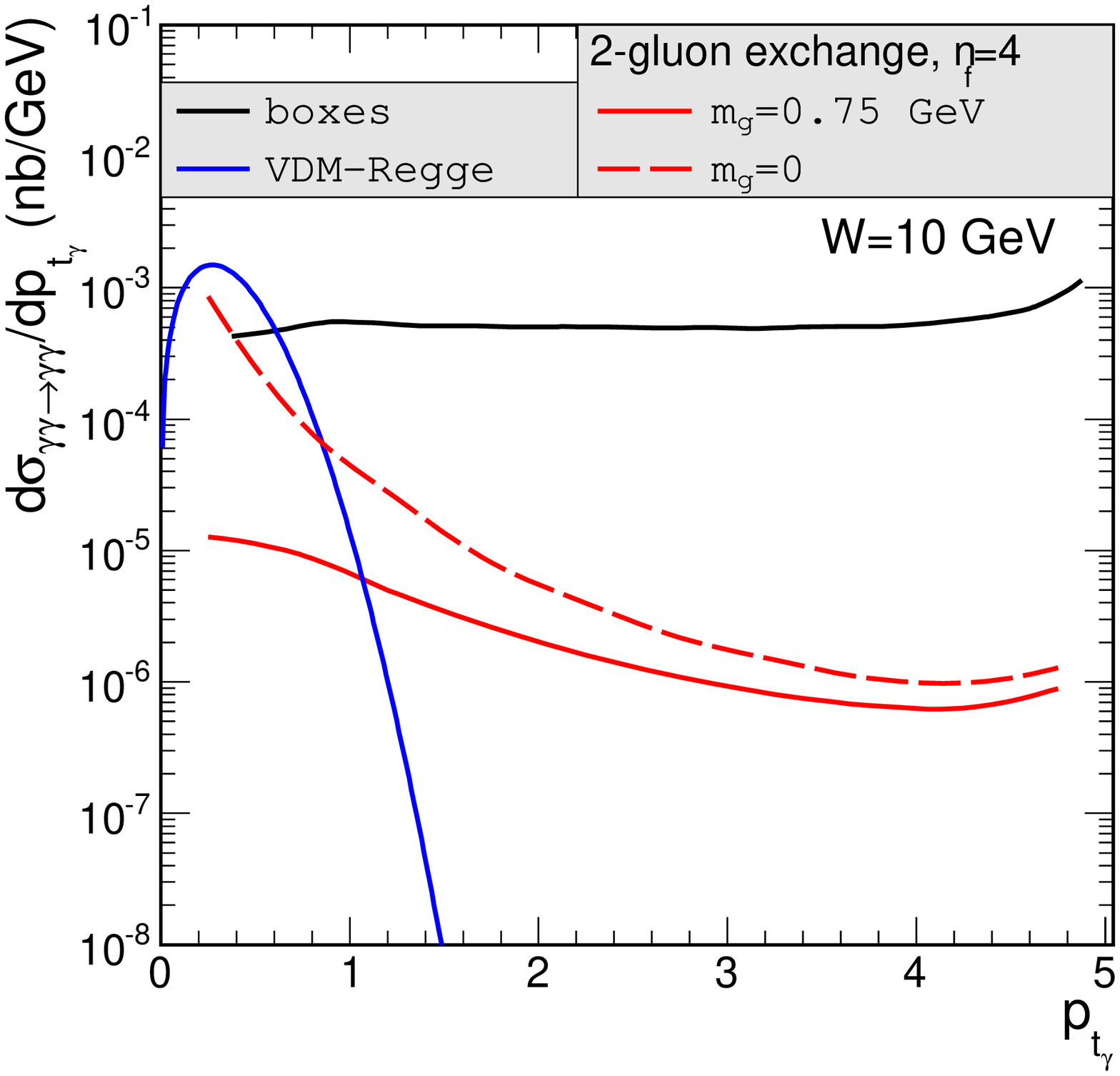}
\includegraphics[width=0.3\textwidth]{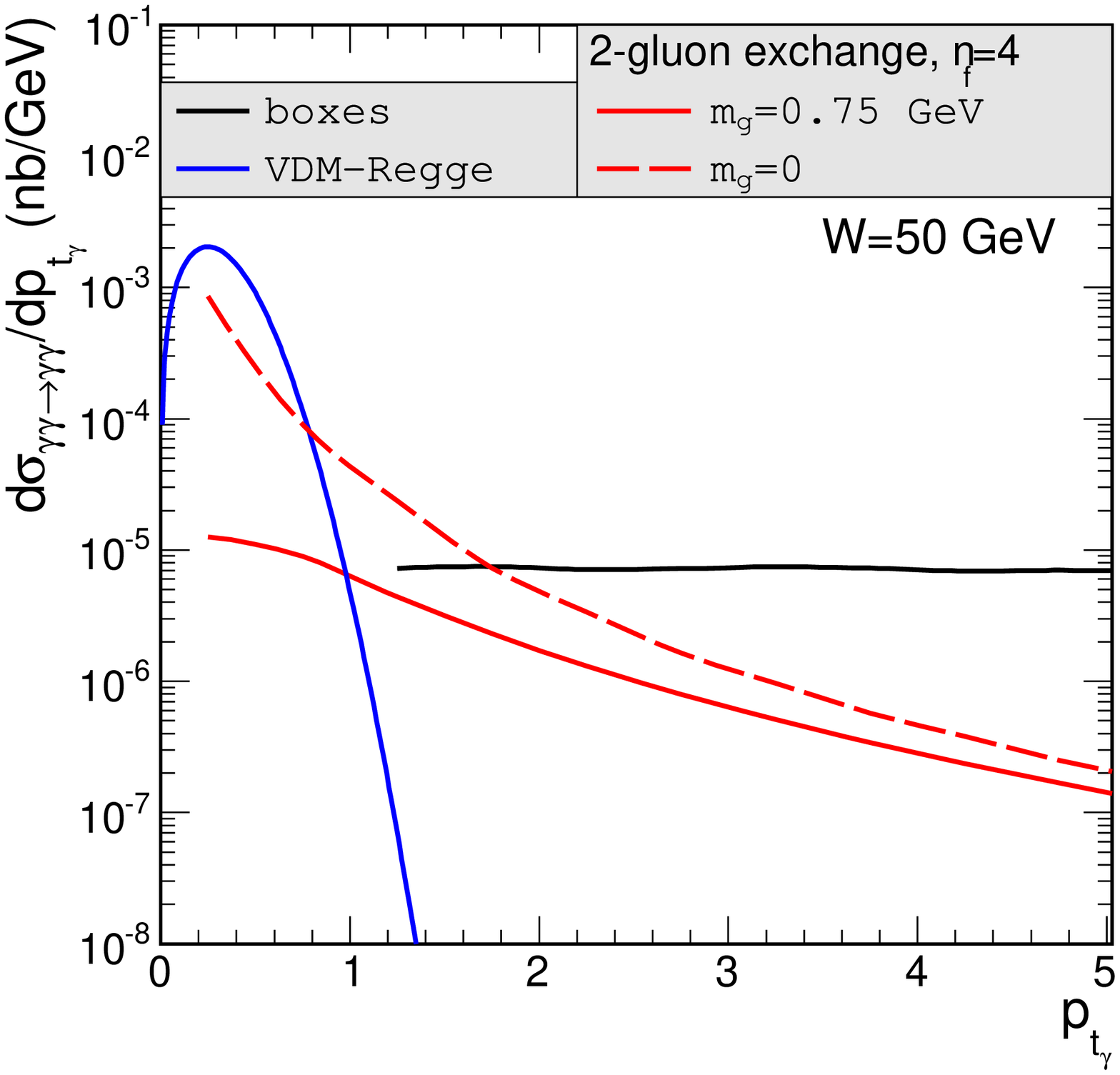}
\includegraphics[width=0.3\textwidth]{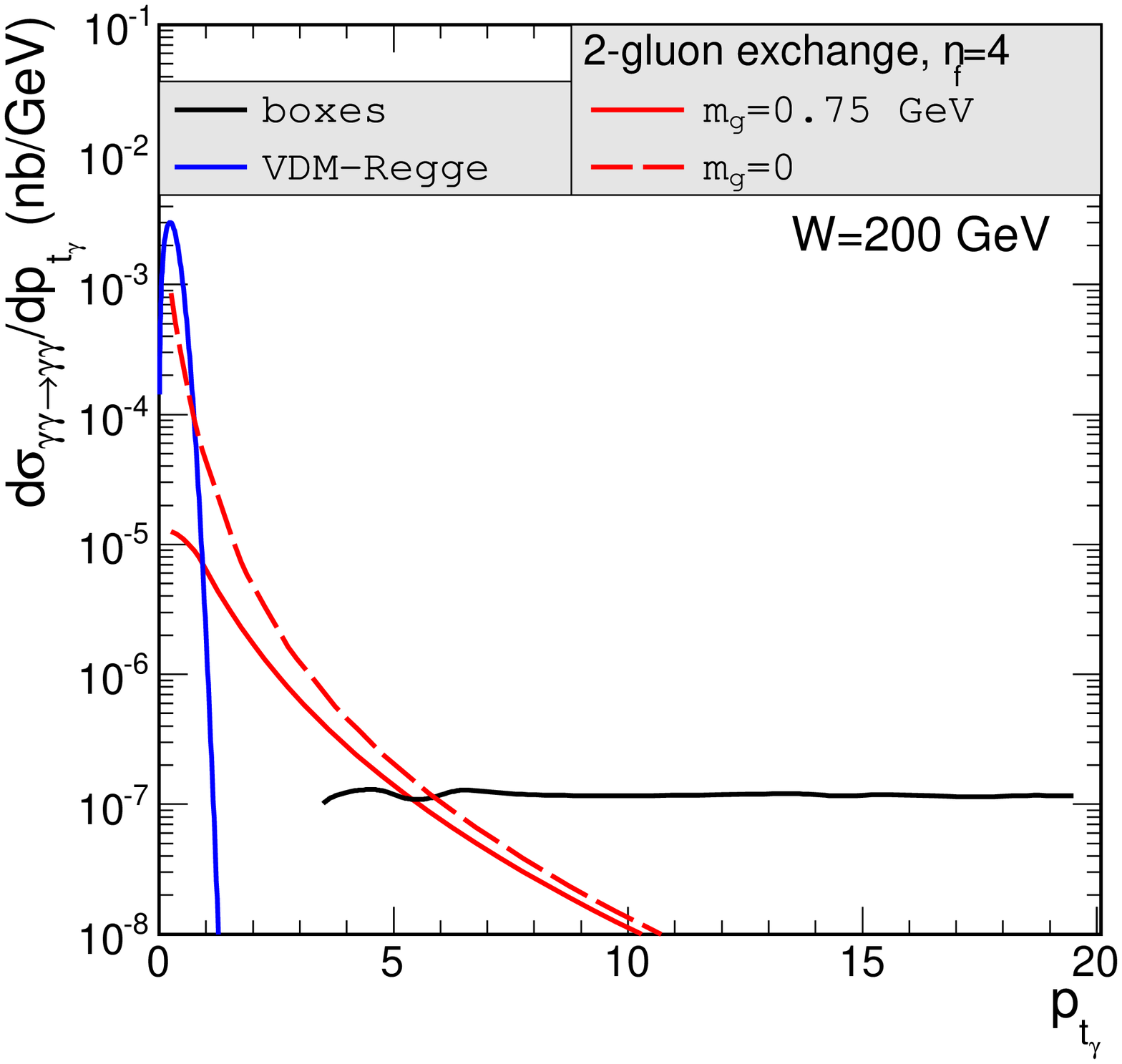}
\caption{Competition of different mechanisms for
transverse momentum dependence of one of outgoing photons for the 
$\gamma \gamma \to \gamma \gamma$ elastic scattering.
Individual contributions are shown separately.
}
\label{fig:dsig_dpt_3processes}
\end{figure}

Can the two-gluon exchange contribution be identified experimentally?
To answer this question in Fig.~\ref{fig:dsig_dpt_3processes}
we show again the three contributions to transverse momentum distribution 
for quite different energies ($W =$ 10, 50, 200 GeV).
The soft VDM-Regge contribution occupies the region of very small
transverse momenta, where it dominates. At low energies the two-gluon exchange
contribution is always smaller than the VDM-Regge and box contributions.
Increasing energy the situation improves 
for observing the influence of the two-gluon exchange mechanism.
At $W =$ 50 GeV there is a
small window of photon transverse momenta 1 GeV $ < p_t < $ 2 GeV where 
its contribution should be seen. At $W =$ 200 GeV the window 
where the two-gluon is
larger than the two other contributions extends now to
1 GeV $ < p_t < $ 5 GeV. 
However, as was already pointed out in \cite{Ginzburg:1985tp} (see also 
\cite{Evanson:1999zb})
potentially a BFKL resummation of large logarithms $\log (\hat s / |\hat t|)$ could lead
to a substantial enhancement of the $\gamma \gamma \to \gamma \gamma$ elastic scattering.
This could be studied in a future.
Clearly the effect could be studied experimentally in a future
photon-photon collider
(the photon-photon collider could be realized at the facility of 
the International $e^+e^-$ Linear Collider (ILC)).
Now we wish to briefly investigate what could be the effect of the two-gluon
exchange mechanism at the LHC both in ultraperipheral heavy-ion collisions
and in exclusive $p p \to p p \gamma \gamma$ processes.

\subsection{$A A \to A A \gamma \gamma$ process}

In the near future ultrarelativistic collisions seems to be the best
place to examine elastic photon-photon collisions \cite{KLS2016}.
In this case the cross section is enhanced by the $Z_1^2 Z_2^2$ factor
compared to the proton-proton collisions, which for lead-lead 
collisions at the LHC ($Z_1 = Z_2 =$ 82) is huge.

As in Ref.\cite{KLS2016} we expect that the distributions in rapidity
or rapidity difference between the two photons could be helpful in 
distinguishing the box and two-gluon exchange contribution.
A lower cut on photon transverse momentum $p_t >$ 1 GeV is necessary to
get rid of the soft region where the VDM-Regge contribution dominates,
as discussed in the previous subsection.

\begin{figure}[!h]
\includegraphics[scale=0.35]{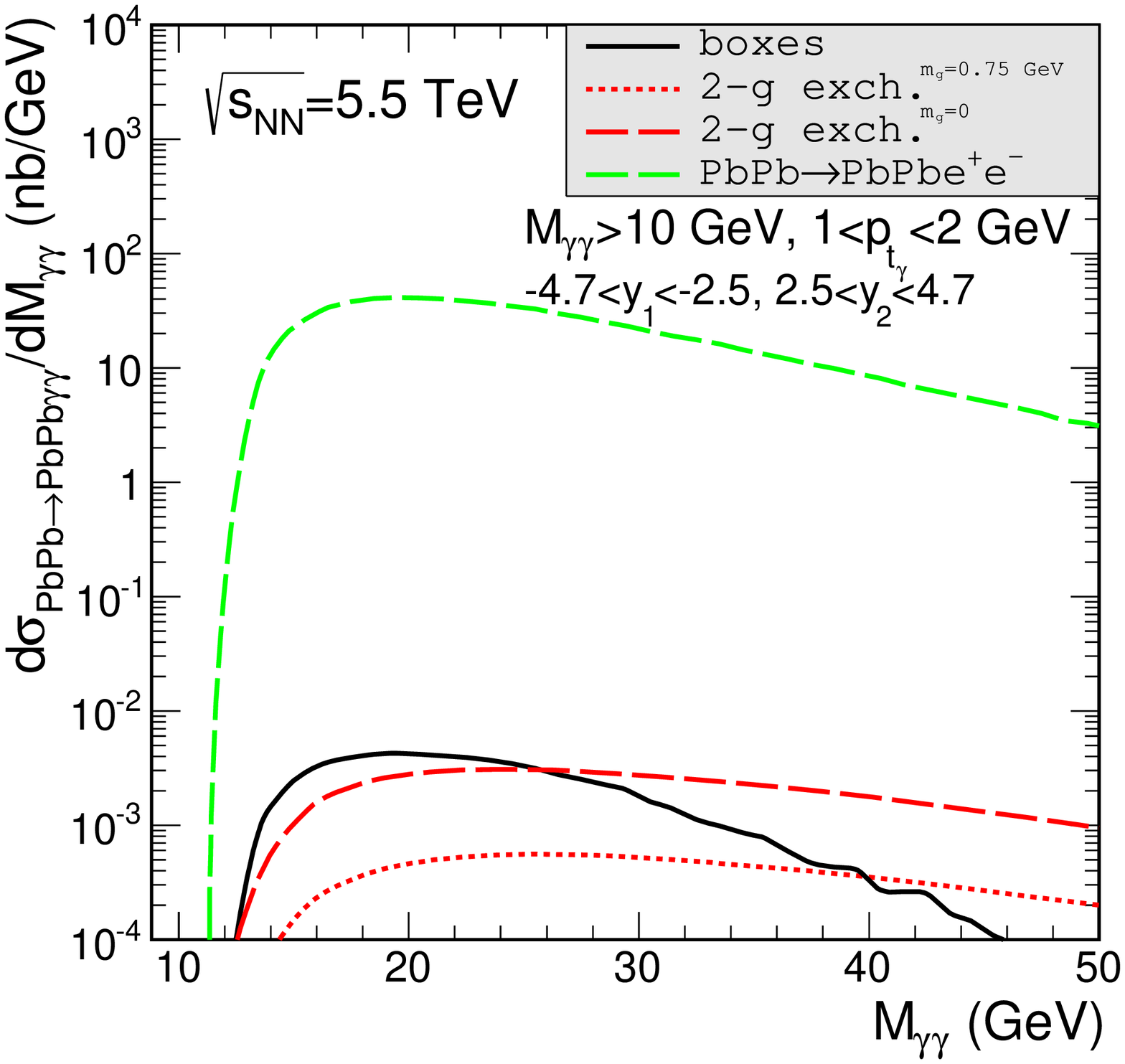}
\includegraphics[scale=0.35]{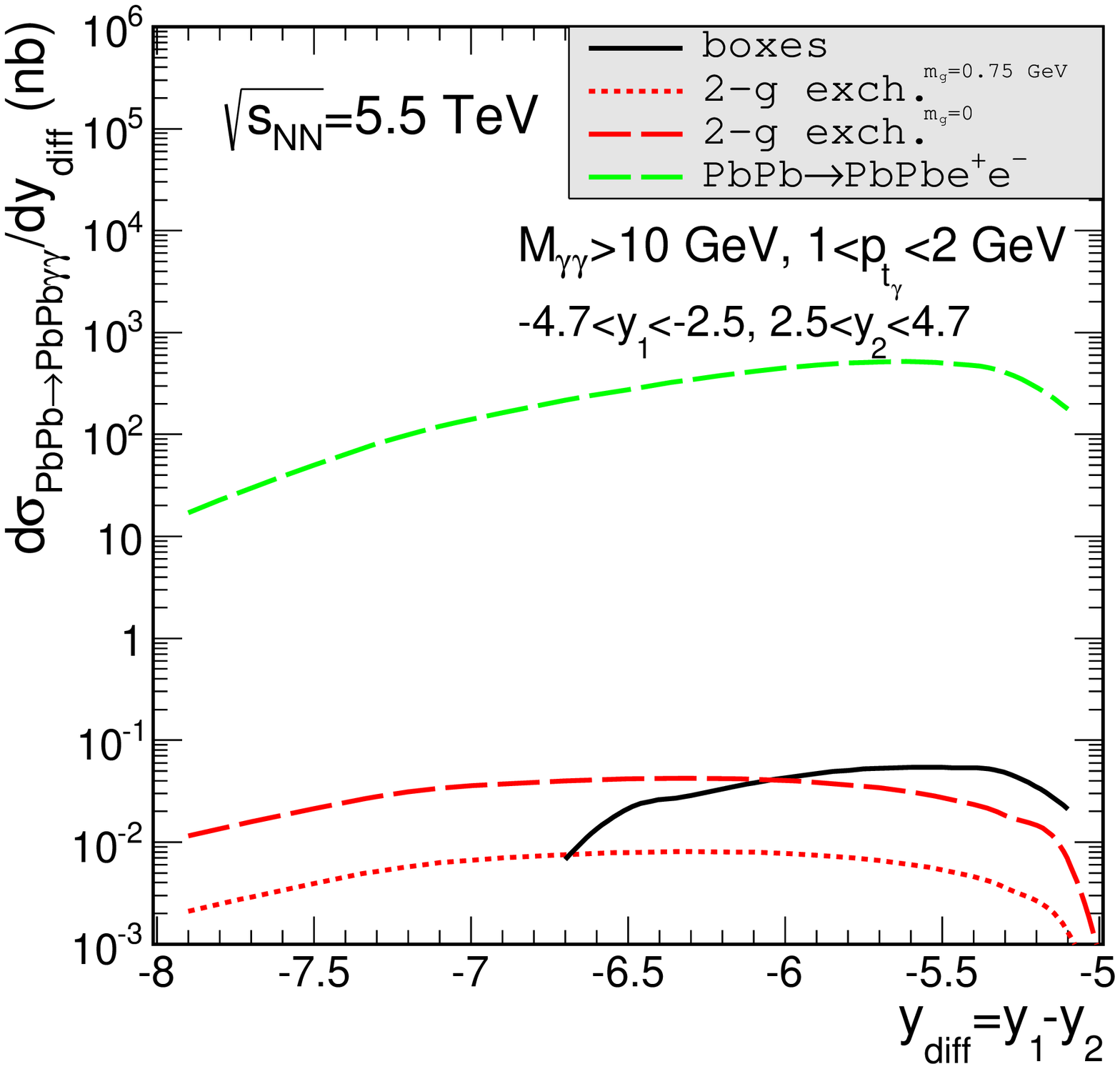}
\caption{Distribution in invariant mass of photons and in rapidity distance between the
two photons for $M_{\gamma\gamma}>10$ GeV,
1 GeV $<p_{t_{\gamma}}<$ 2 GeV and $-4.7<y_1<-2.5$, $2.5<y_2<4.7$. 
In addition, we show (top dashed, green line) a similar distribution for $AA \to AA e^+e^-$.
}
\label{fig:dsig_dW_dydiff_nucl}
\end{figure}

We see that the bigger distance between photons, 
the larger two-gluon to box contribution ratio is.
Therefore we consider also a possibility to observe
photons with forward calorimeters (FCALs).
In Fig.~\ref{fig:dsig_dW_dydiff_nucl}
we show differential distribution as a function
of $M_{\gamma\gamma}$ and $y_{diff}=y_1-y_2$.
The results are shown both for box and two-gluon exchange mechanisms. 
For comparison we also show contribution
which comes from a $\gamma\gamma \to e^+e^-$ subprocess.
We emphasise that this subprocess is a (reduceable) background to the 
light-by-light scattering.

\begin{figure}[!h]
\includegraphics[width=5cm]{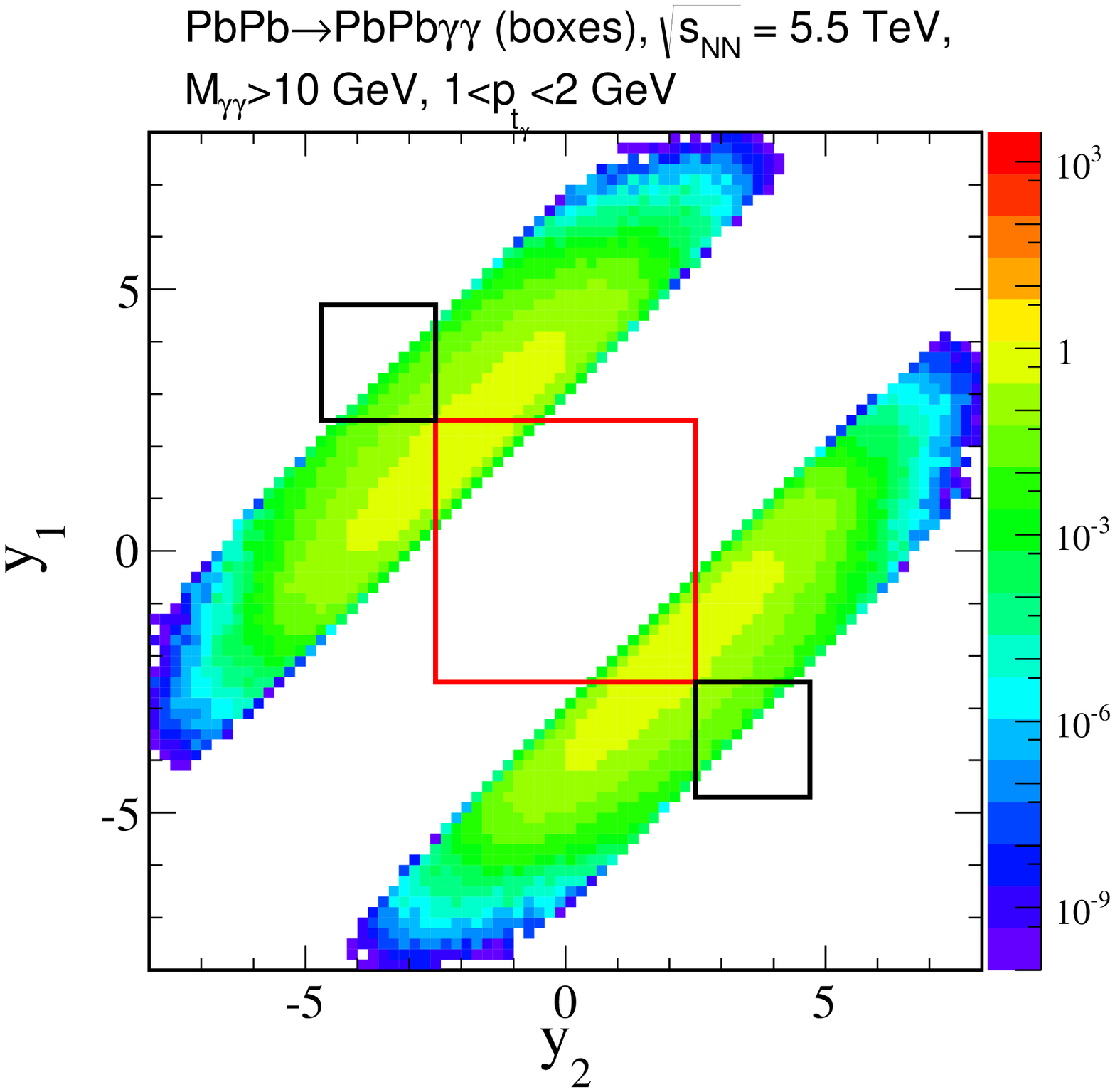}
\includegraphics[width=5cm]{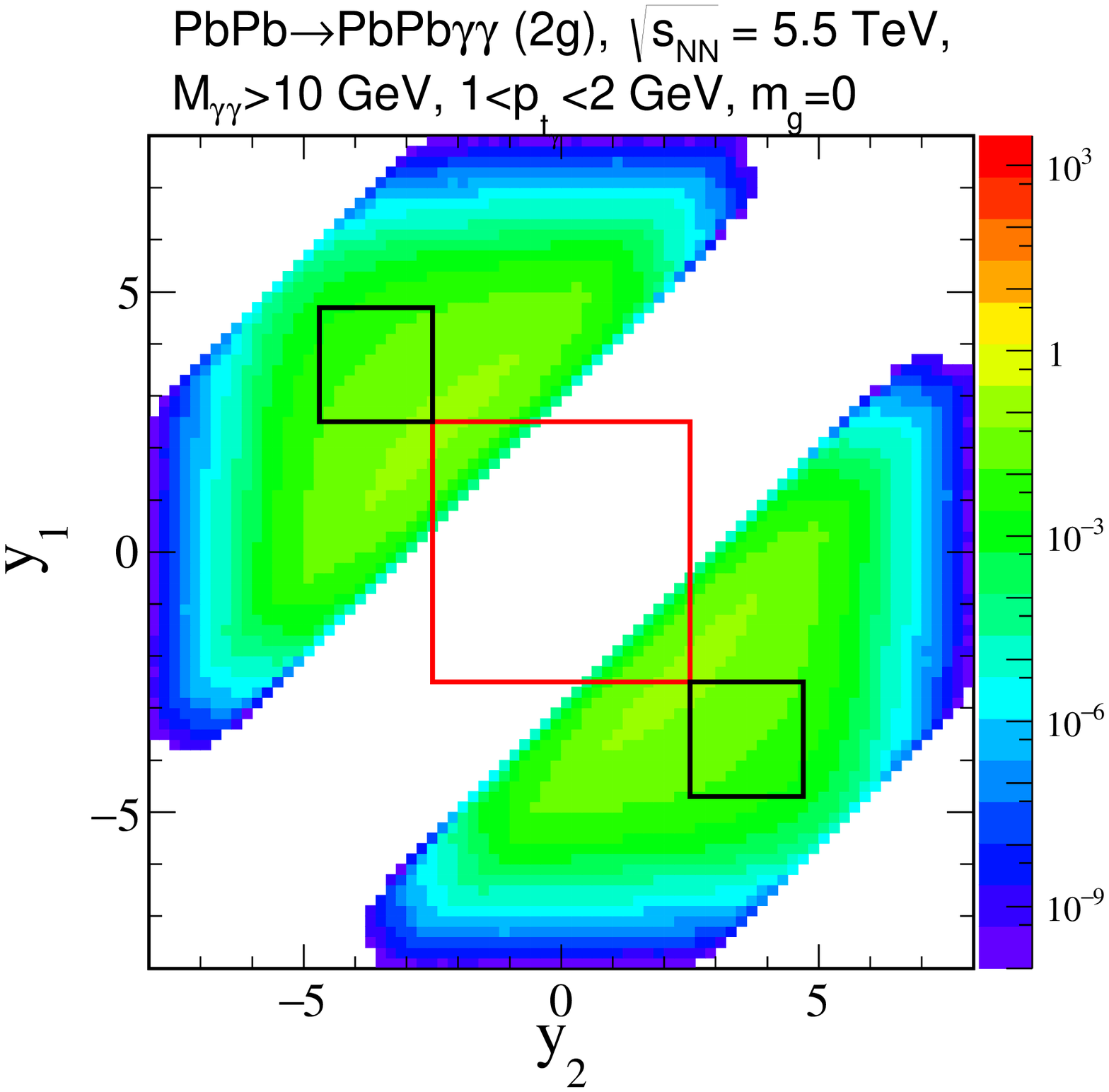}
\includegraphics[width=5cm]{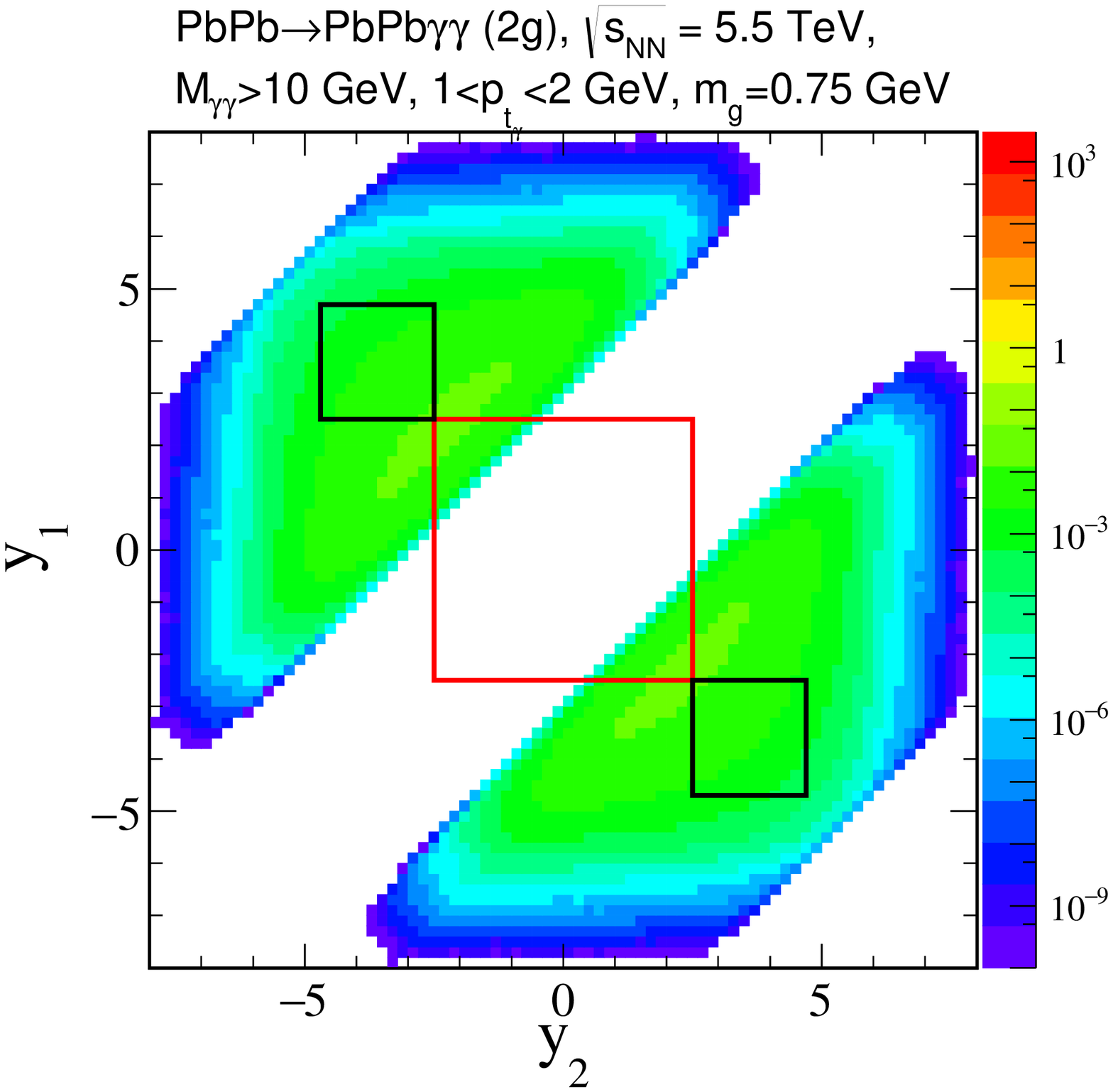}
\includegraphics[width=5cm]{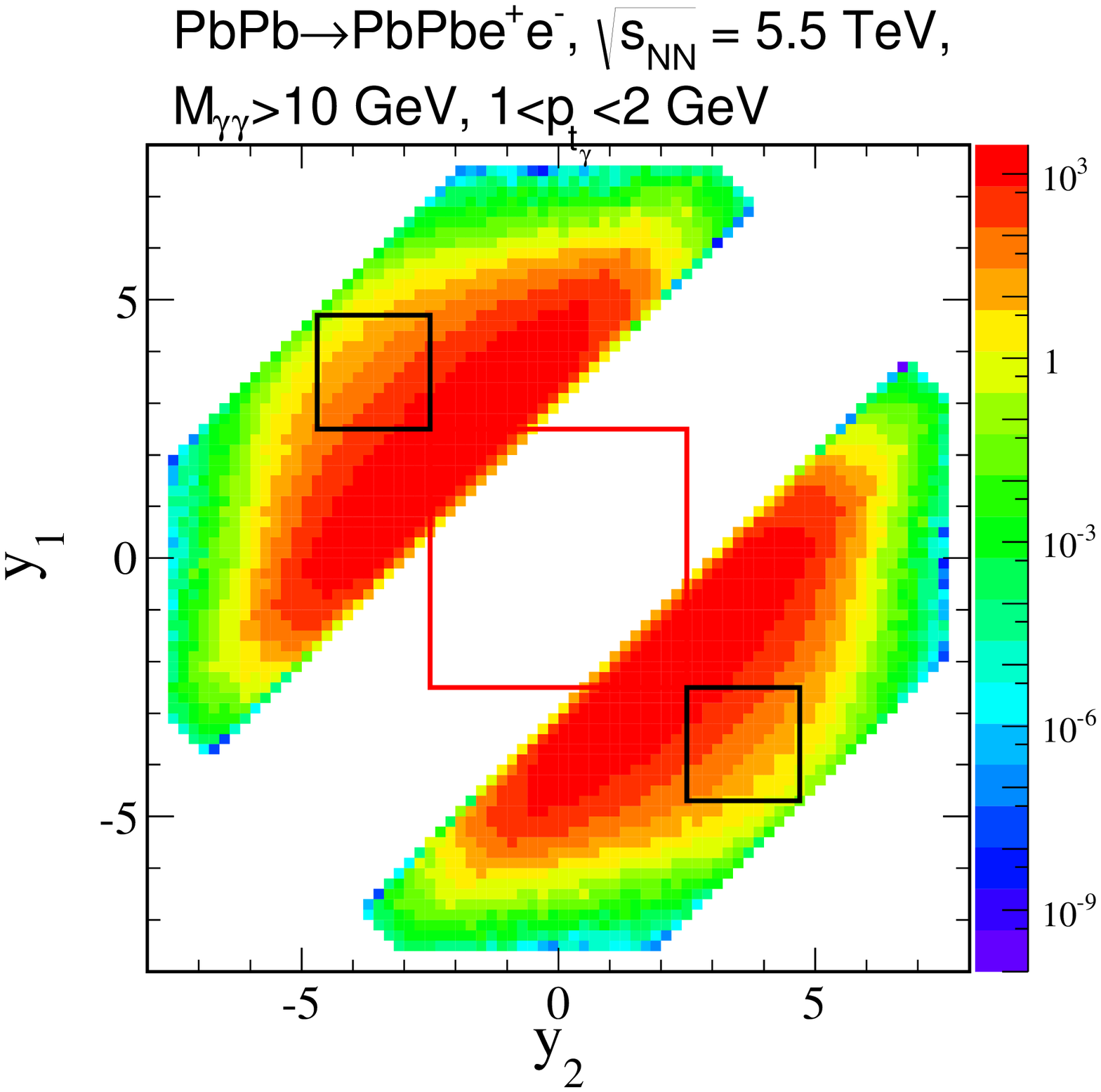}
\caption{Two-dimensional distributions in rapidities of the produced photons
for the box mechanism, the two-gluon exchange mechanism and for the 
$AA \to AA e^+e^-$ process. Cuts on $M_{\gamma \gamma}$ and photon transverse momenta 
are specified in the figure legend.
}
\label{fig:dsig_dy1dy2_nucl}
\end{figure}

In Fig.~\ref{fig:dsig_dy1dy2_nucl} we present 
$d\sigma/dy_1dy_2$ map for boxes, two-gluon exchange mechanism 
(for $m_g=0$ and $m_g=0.75$  GeV) 
and for comparison a result for $PbPb \to PbPb e^+e^-$. 
We denote the coverage of the main detector 
($-2.5<y_{1/2}<2.5$ - red square)
and two smaller squares which represent situations when
one photon is in one-side forward calorimeter and the second photon 
is in the second-side forward calorimeter.

\subsection{$p p \to p p \gamma \gamma$ process}

\begin{figure}[!h]
\includegraphics[width=5cm]{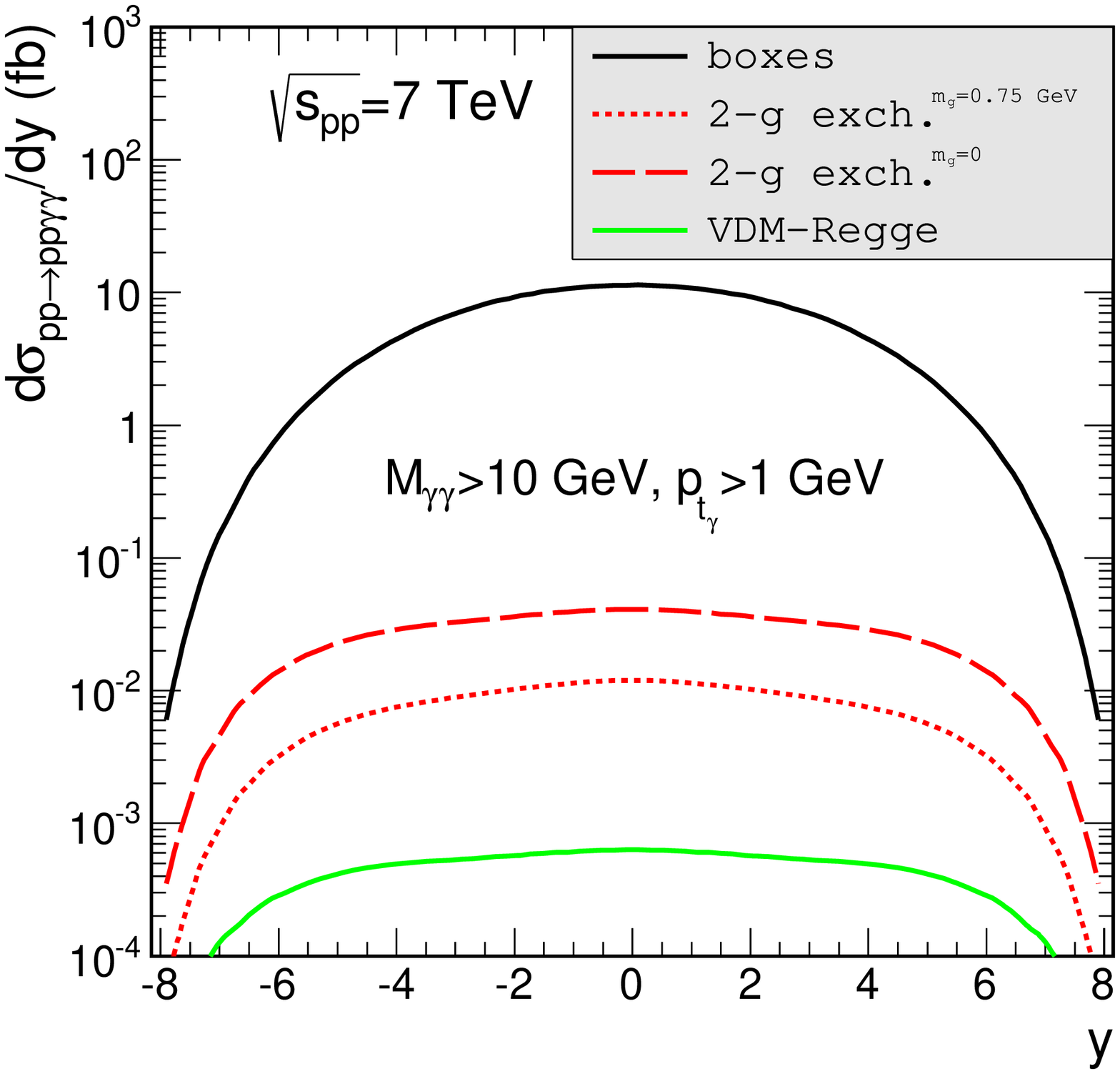}
\includegraphics[width=5cm]{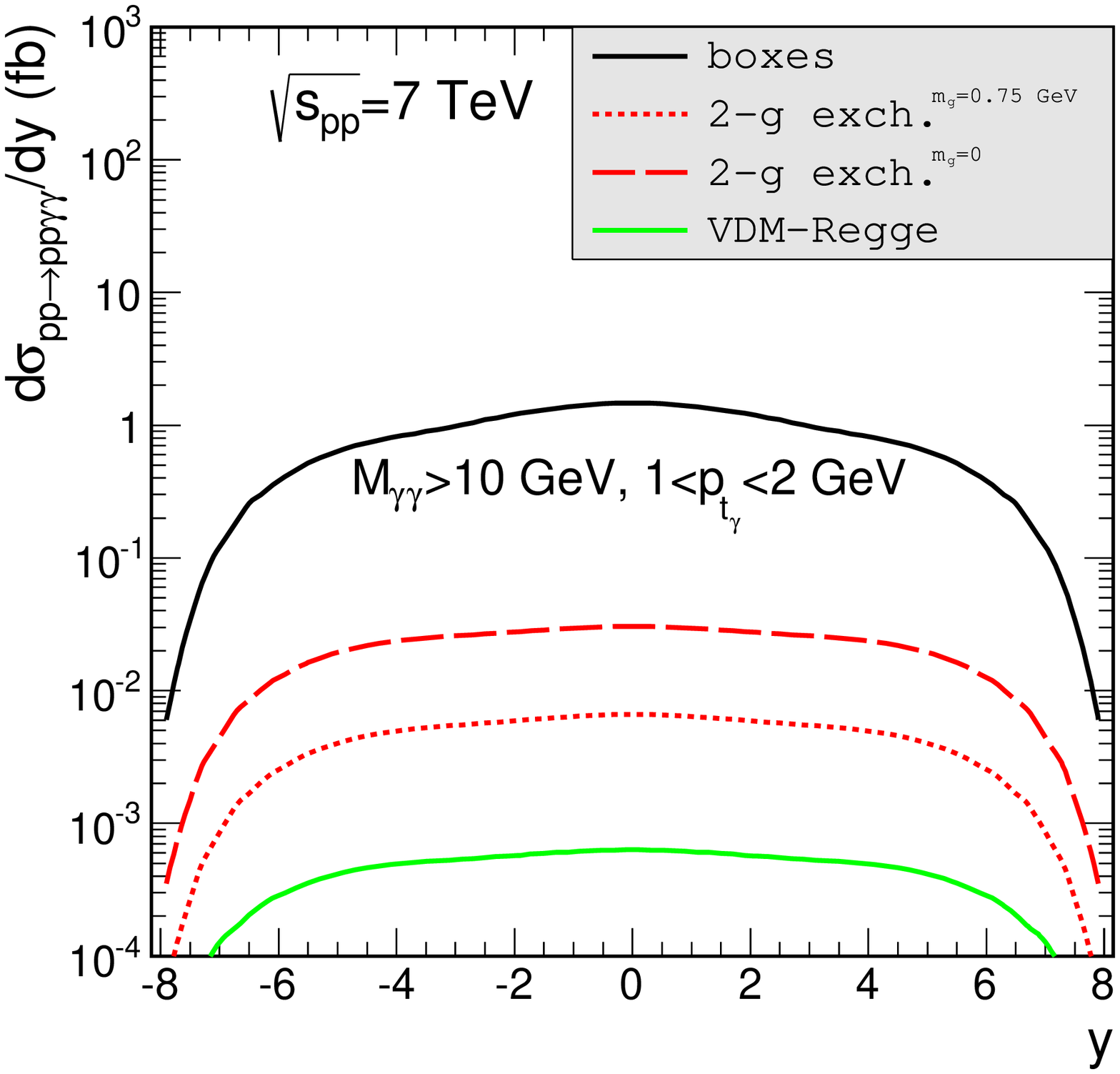}
\includegraphics[width=5cm]{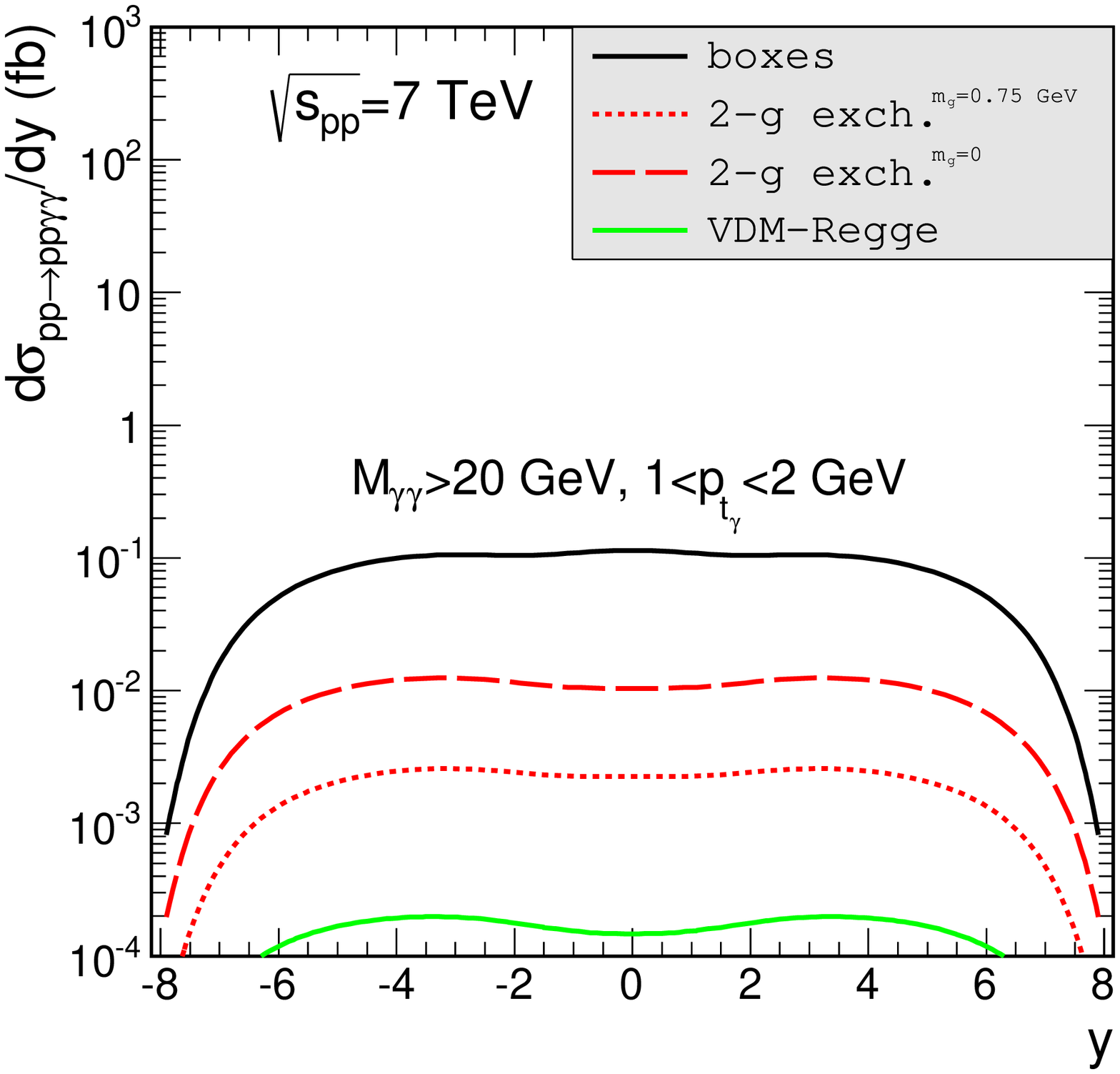}
\caption{Distribution in rapidity of one of the photons
for different cuts specified in the figure legend.
}
\label{fig:dsig_dy1}
\end{figure}

The $p p \to p p \gamma \gamma$ reaction is an alternative for the 
$AA \to AA \gamma \gamma$ studies.
In the following we discuss first results for the $p p \to p p \gamma
\gamma$ reaction.
In Fig.~\ref{fig:dsig_dy1} we show distribution in rapidity of one of 
the photons. The results are for different cuts on $M_{\gamma\gamma}$
and transverse momentum of each of the photons.
The results for two-gluon exchange contribution are shown with $m_g$ = 0
(upper curve) and $m_g$ = 0.75 GeV (lower curve). 
Even with the restrictive cuts, the two-gluon exchange contribution
is less than 10 \% of that for the boxes. The VDM-Regge contribution
is very small (negligible) as we have imposed lower cut on photon 
transverse momentum $p_t >$ 1 GeV. As discussed in 
Ref.\cite{KLS2016} the VDM-Regge contribution is very soft, concentrated
dominantly for $p_t <$ 1 GeV.

\begin{figure}[!h]
\includegraphics[width=5cm]{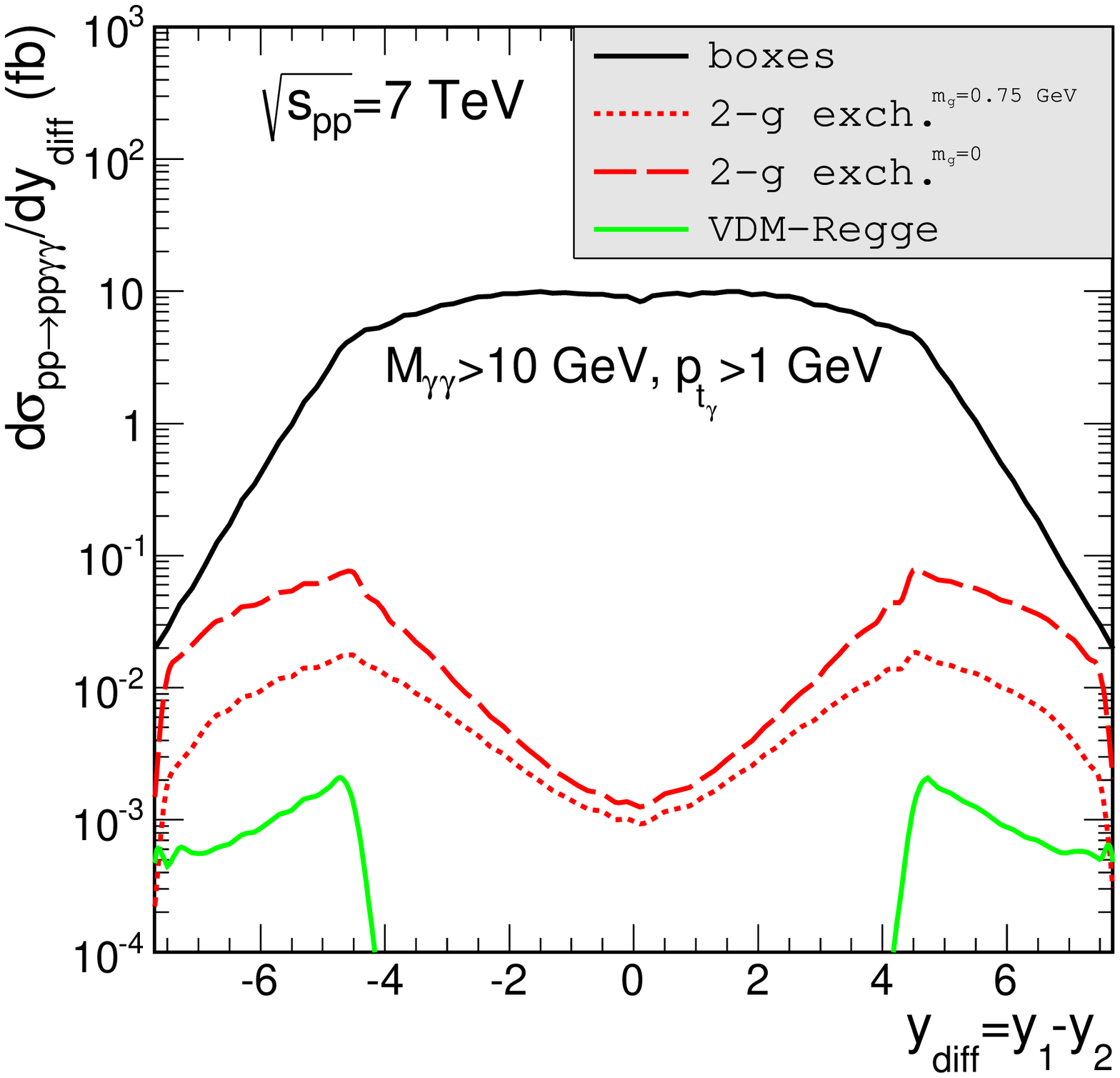}
\includegraphics[width=5cm]{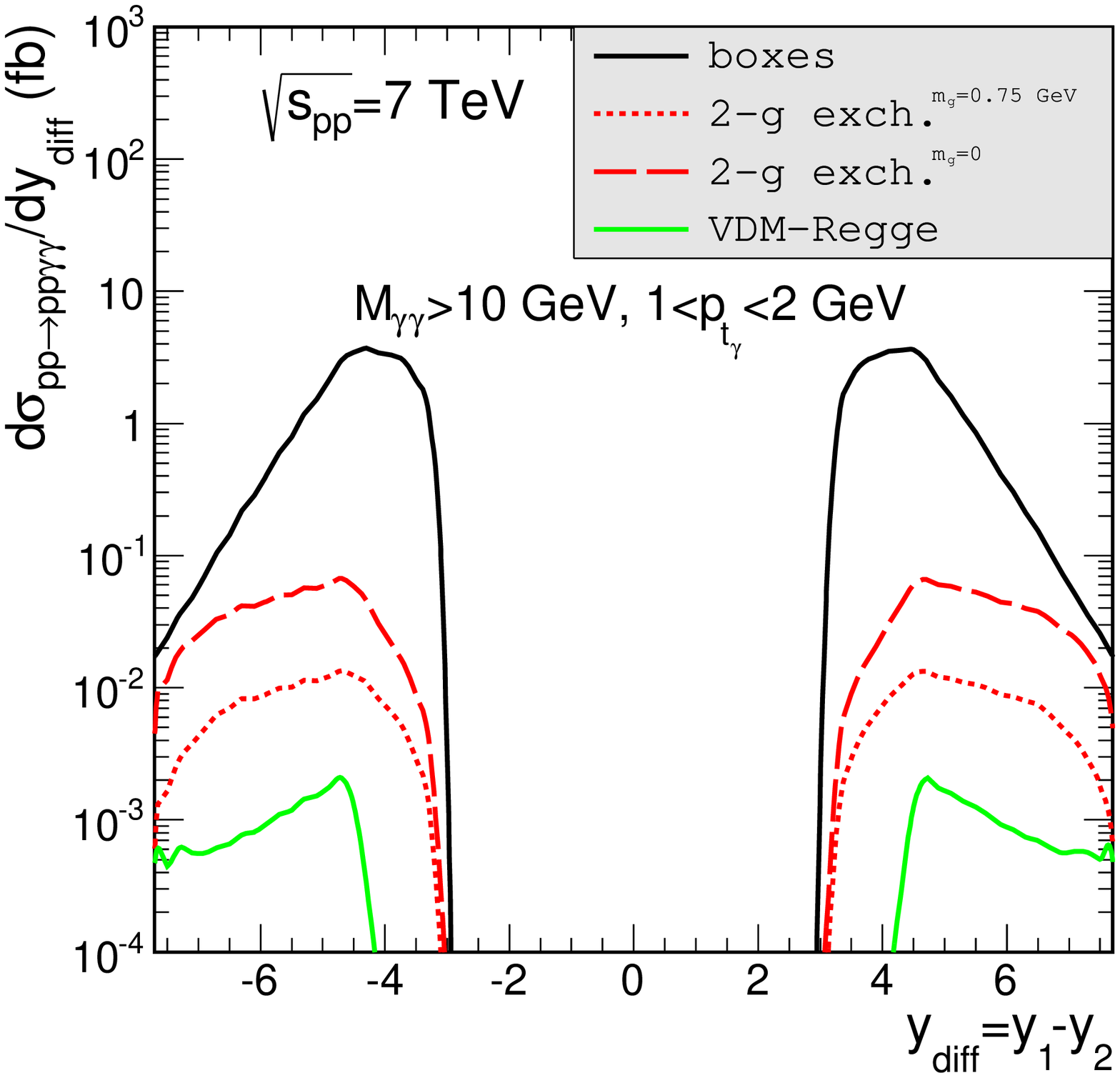}
\includegraphics[width=5cm]{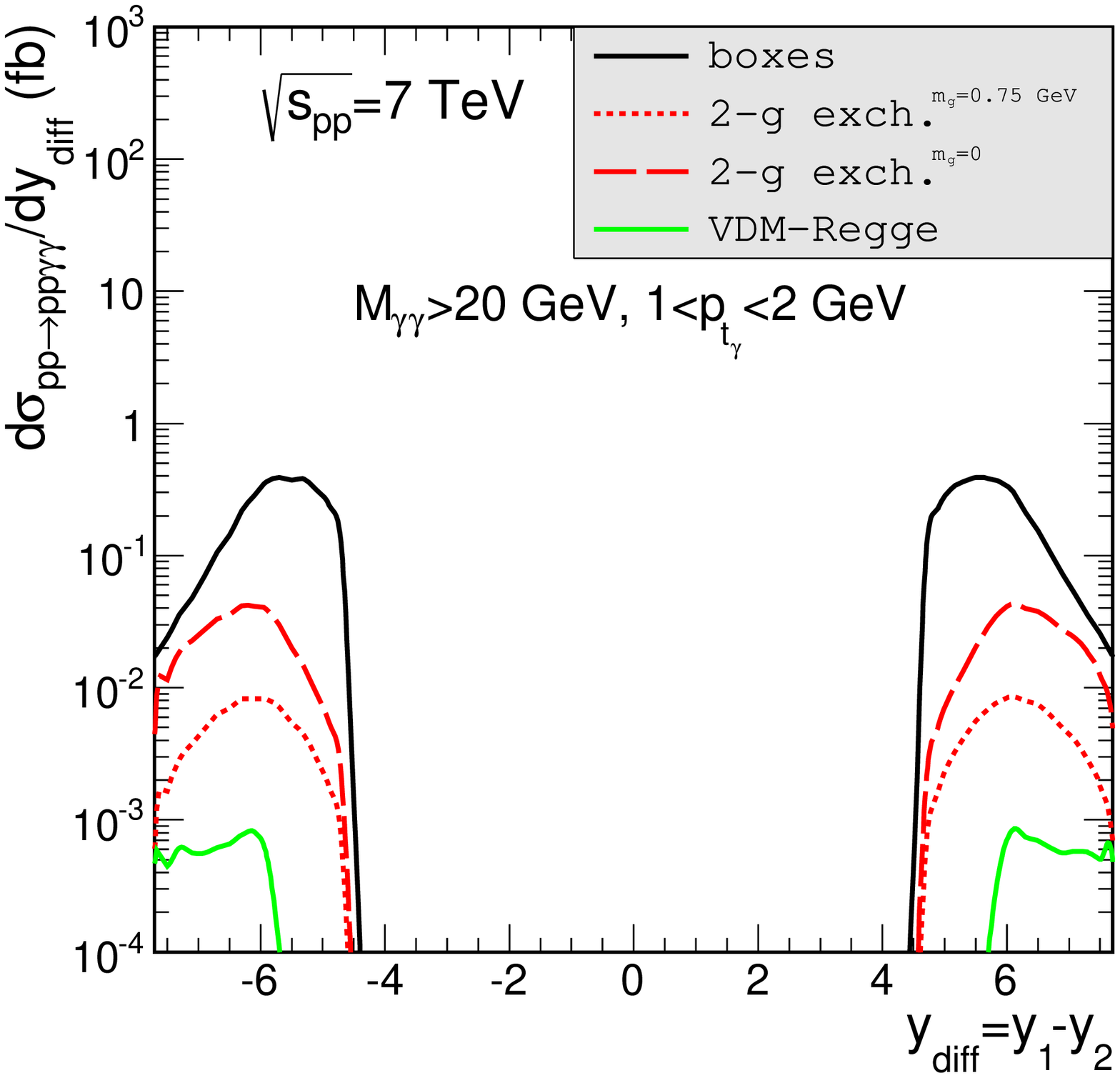}
\caption{Distribution in rapidity distance between the
two photons for different cuts specified in the figure legend.
No cuts on photon rapidities are applied here.
}
\label{fig:dsig_dydiff}
\end{figure}

The distribution in rapidity distance between both photons seems
more promising (see Fig.~\ref{fig:dsig_dydiff}). 
Increasing the lower cut on $M_{\gamma \gamma}$
and limiting to a narrow window in photon transverse momenta
improves the relative amount of the two-gluon exchange contribution. 

\begin{figure}[!h]
\includegraphics[scale=0.25]{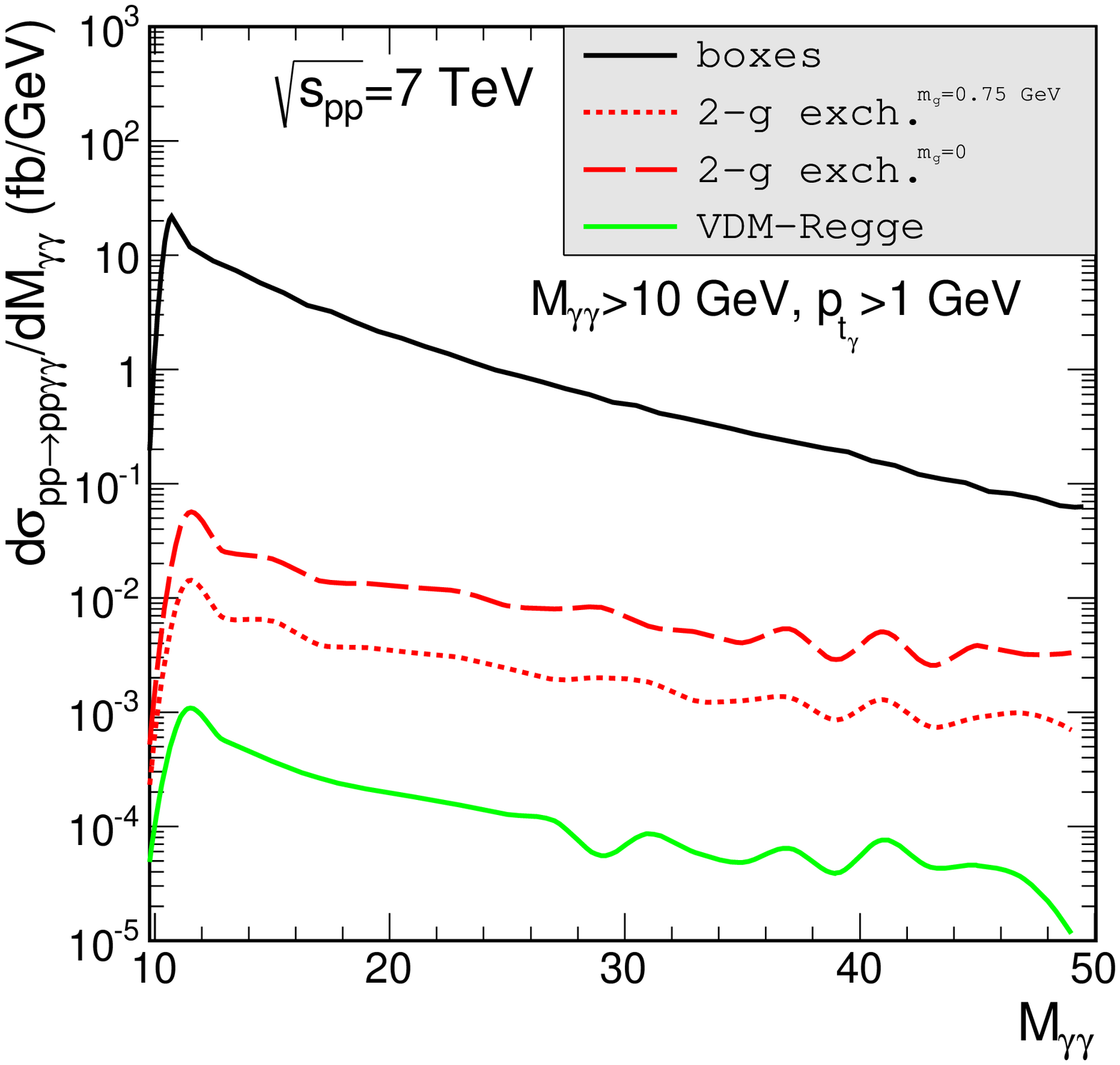}
\includegraphics[scale=0.25]{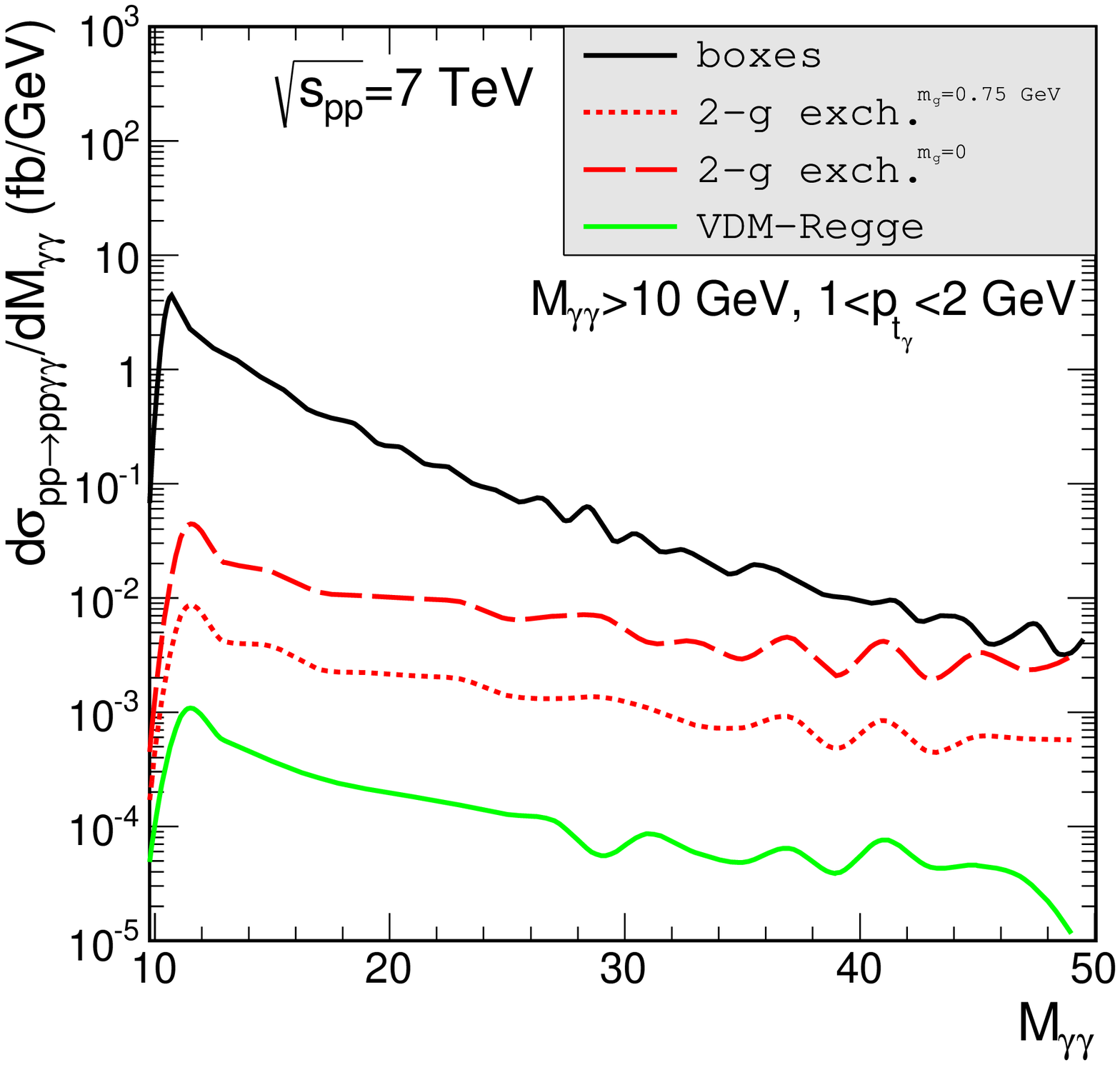}
\includegraphics[scale=0.25]{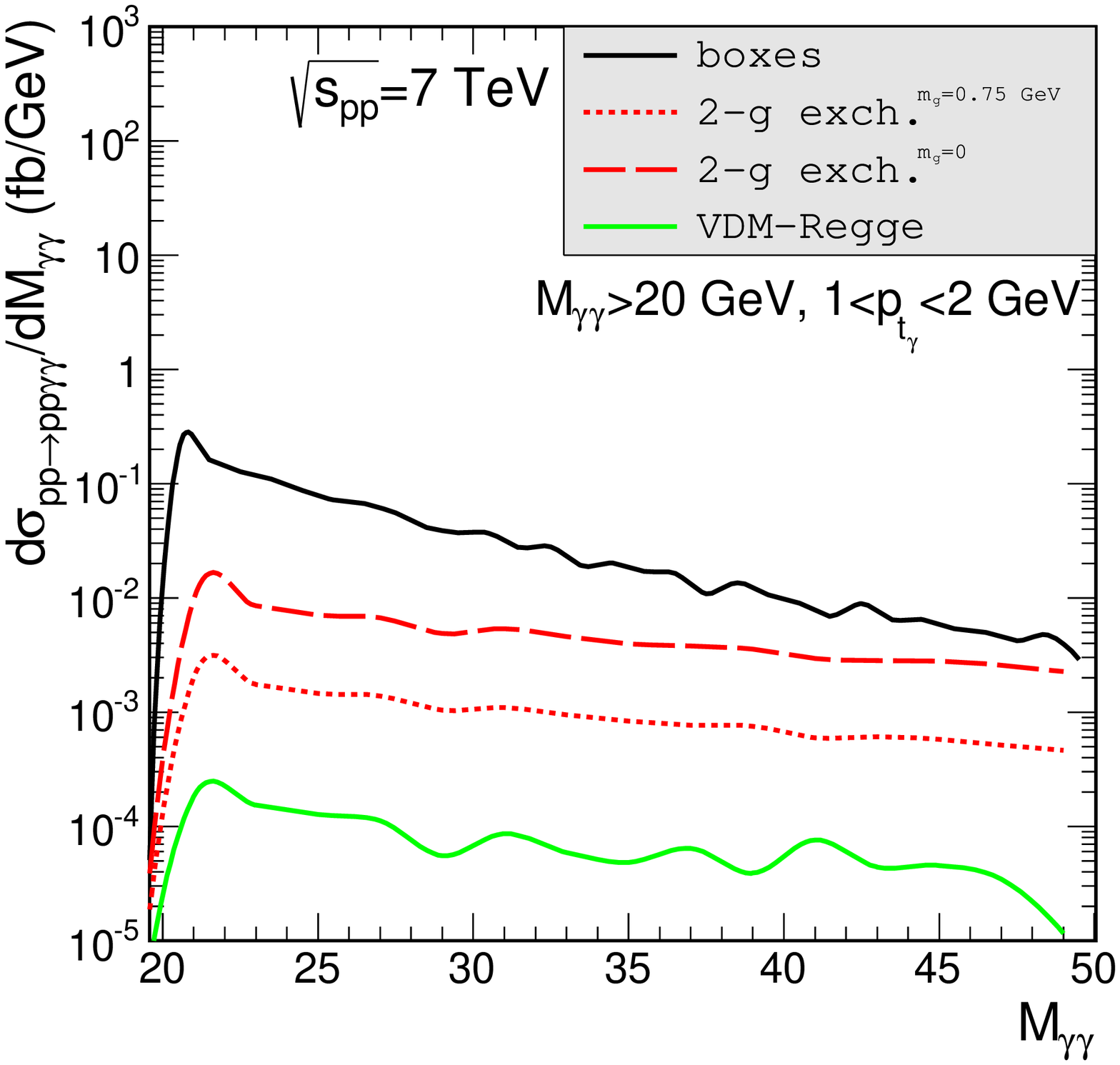}
\caption{Distribution in invariant mass of the produced photons
for different cuts specified in the figure legend.
No cuts on photon rapidities are applied here.
}
\label{fig:dsig_dW}
\end{figure}

The distribution in the diphoton invariant mass is shown in Fig.
\ref{fig:dsig_dW}. The two-gluon distribution starts to dominante
over the box contribution only above $M_{\gamma \gamma} >$ 50 GeV
for 1~GeV $<p_t<$ 5 GeV.
However, the cross section in this region is rather small.

\begin{figure}[!h]
\includegraphics[scale=0.35]{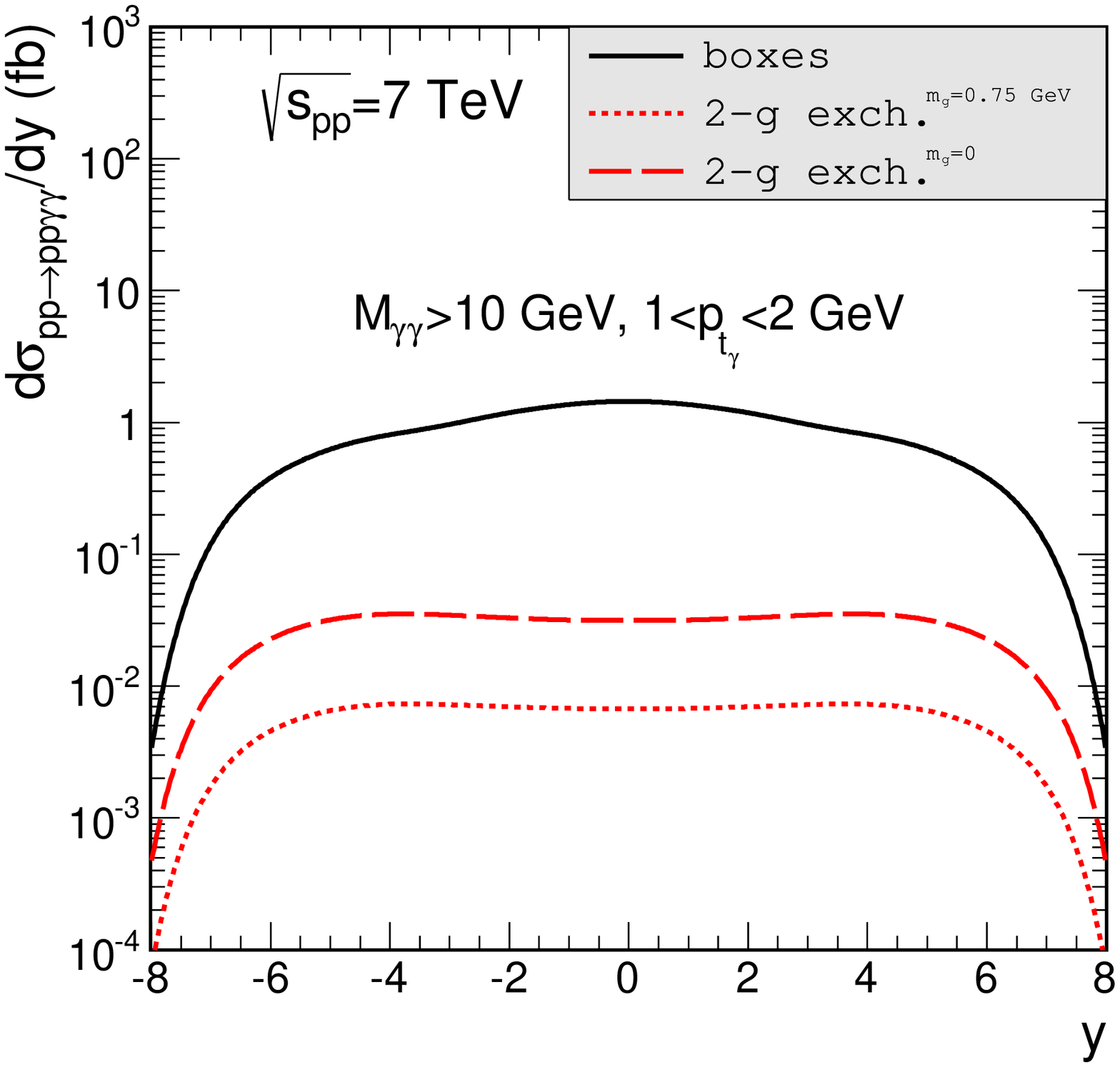}
\includegraphics[scale=0.35]{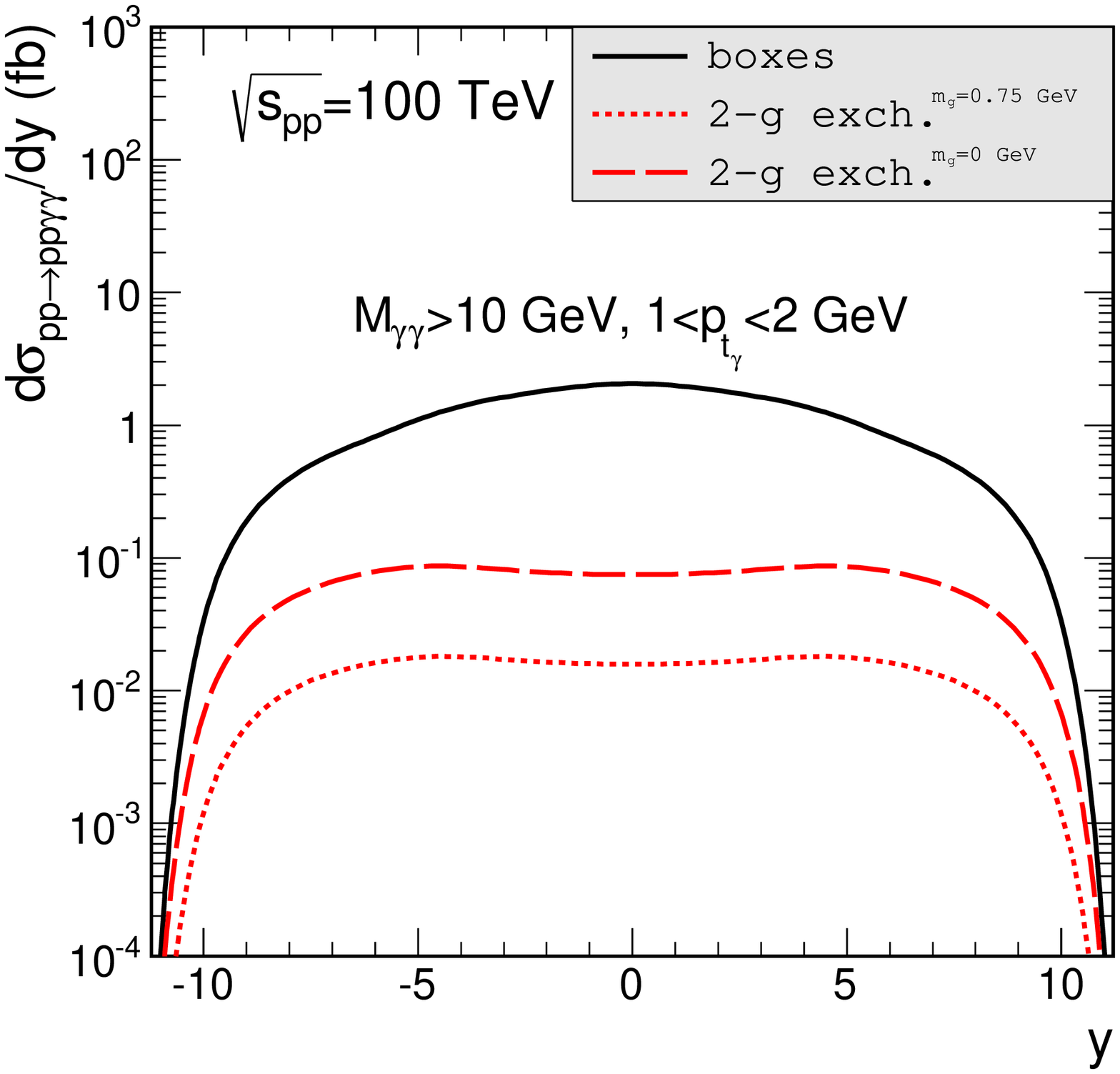}
\caption{Distribution in rapidity of the produced photons
for $\sqrt{s}$ = 7 TeV (LHC) and $\sqrt{s}$ = 100 TeV (FCC) 
for cuts on photon transverse momenta specified in the figure legend.
No cuts on photon rapidities are applied here.
}
\label{fig:dsig_dy1_LHC_vs_FCC}
\end{figure}

\begin{figure}[!h]
\includegraphics[scale=0.35]{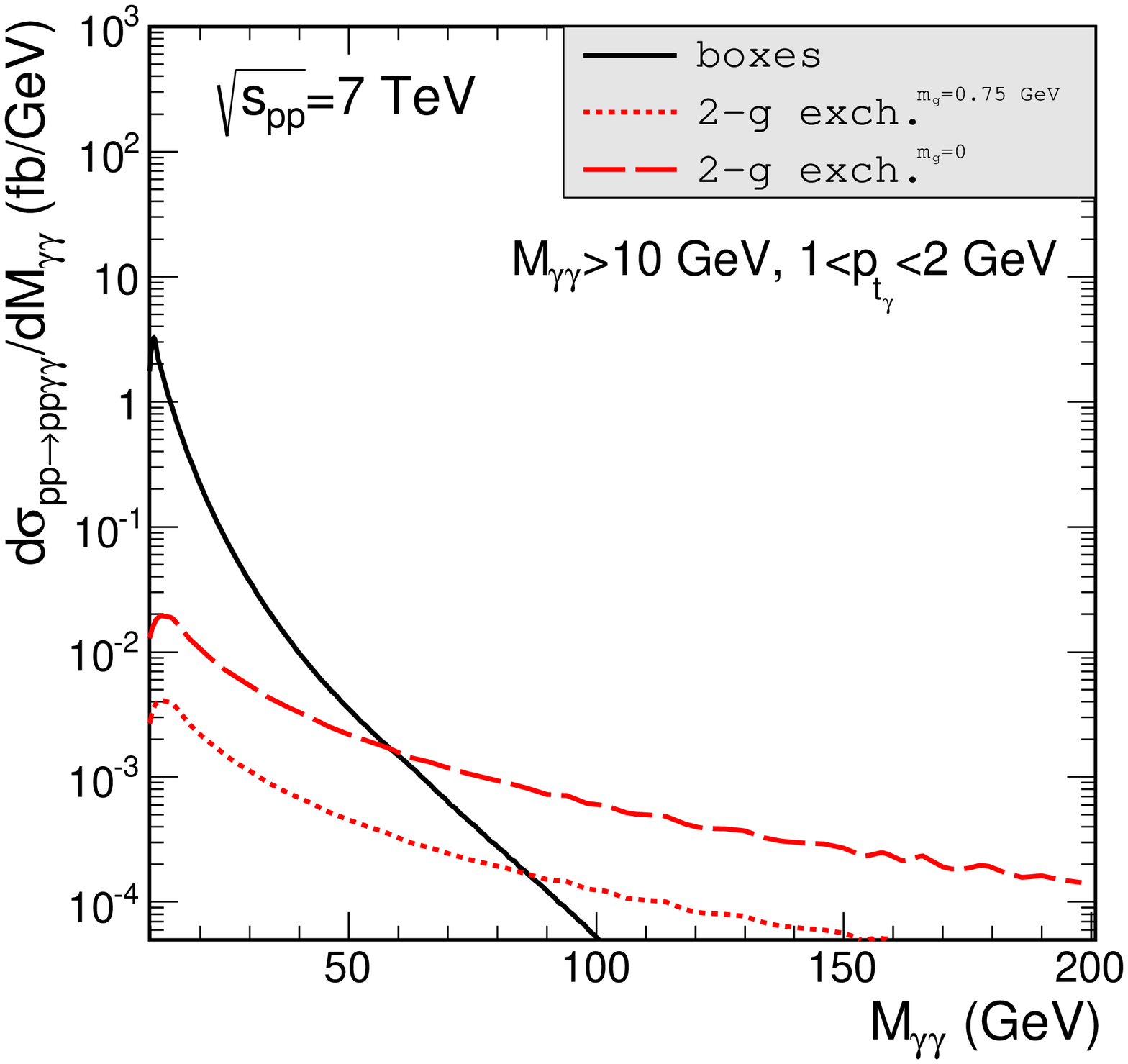}
\includegraphics[scale=0.35]{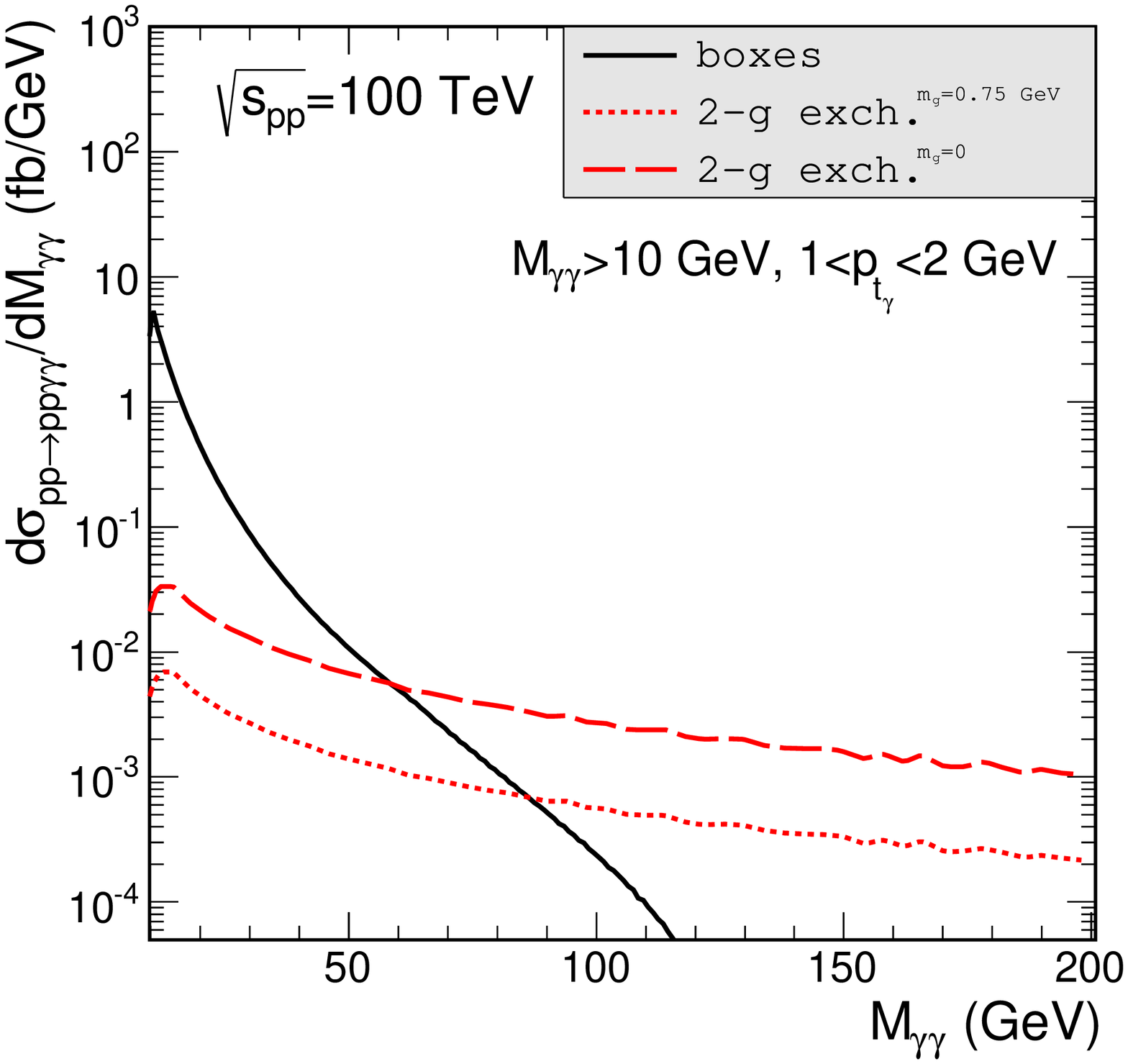}
\caption{Distribution in invariant mass of the produced photons
for $\sqrt{s}$ = 7 TeV (LHC) and $\sqrt{s}$ = 100 TeV (FCC) 
for cuts on photon transverse momenta specified in the figure legend.
No cuts on photon rapidities are applied here.
}
\label{fig:dsig_dW_LHC_vs_FCC}
\end{figure}

How the situation could change for larger collision energies is shown in
Fig.~\ref{fig:dsig_dy1_LHC_vs_FCC} and Fig.\ref{fig:dsig_dW_LHC_vs_FCC}.
The situation for the two presented distributions is rather similar
for the LHC and Future Circular Collider (FCC)\footnote{The box contribution 
to the $pp$, $pA$ and $AA$ cross sections at FCC was estimated in \cite{Enterria_FCC}.}. 
One advantage of larger collision energies are
slightly larger cross sections. However, the dominance of the two-gluon
exchange over the box contribution takes place more or less at 
the same diphoton invariant masses.

Below in Fig.~\ref{fig:dsig_dy1dy2} we show two-dimensional distributions
in rapidities of photons produced in the $p p \to p p \gamma \gamma$
reaction. In this calculation we have assumed a cut on 
$M_{\gamma \gamma} >$ 10 GeV and selected a narrow window on photon transverse
momenta 1 GeV $< p_t <$ 2 GeV. The two-gluon exchange contribution
starts to be larger only in very corner of the phase space
when $|y_1 - y_2|$ is very large.
We have marked the rapidity span of the main (CMS or ATLAS) detector 
(red central square) as well as for forward
calorimeters (black smaller squares). 
It would be interesting to analyze whether the use of forward
calorimeters could be possible in this context. Then the observation 
of one photon in one-side calorimeter and the second photon in 
the second-side calorimeter could help in 
observing the two-gluon exchange contribution. 
It is rather difficult to distinguish photons and electrons with 
the help of the calorimeters.
At such a big rapidity distances (5 $< y_{diff}<$ 9.4) the exclusive 
dielectron contribution \cite{Kubasiak-Szczurek} could be smaller.

\begin{figure}[!h]
\includegraphics[width=5cm]{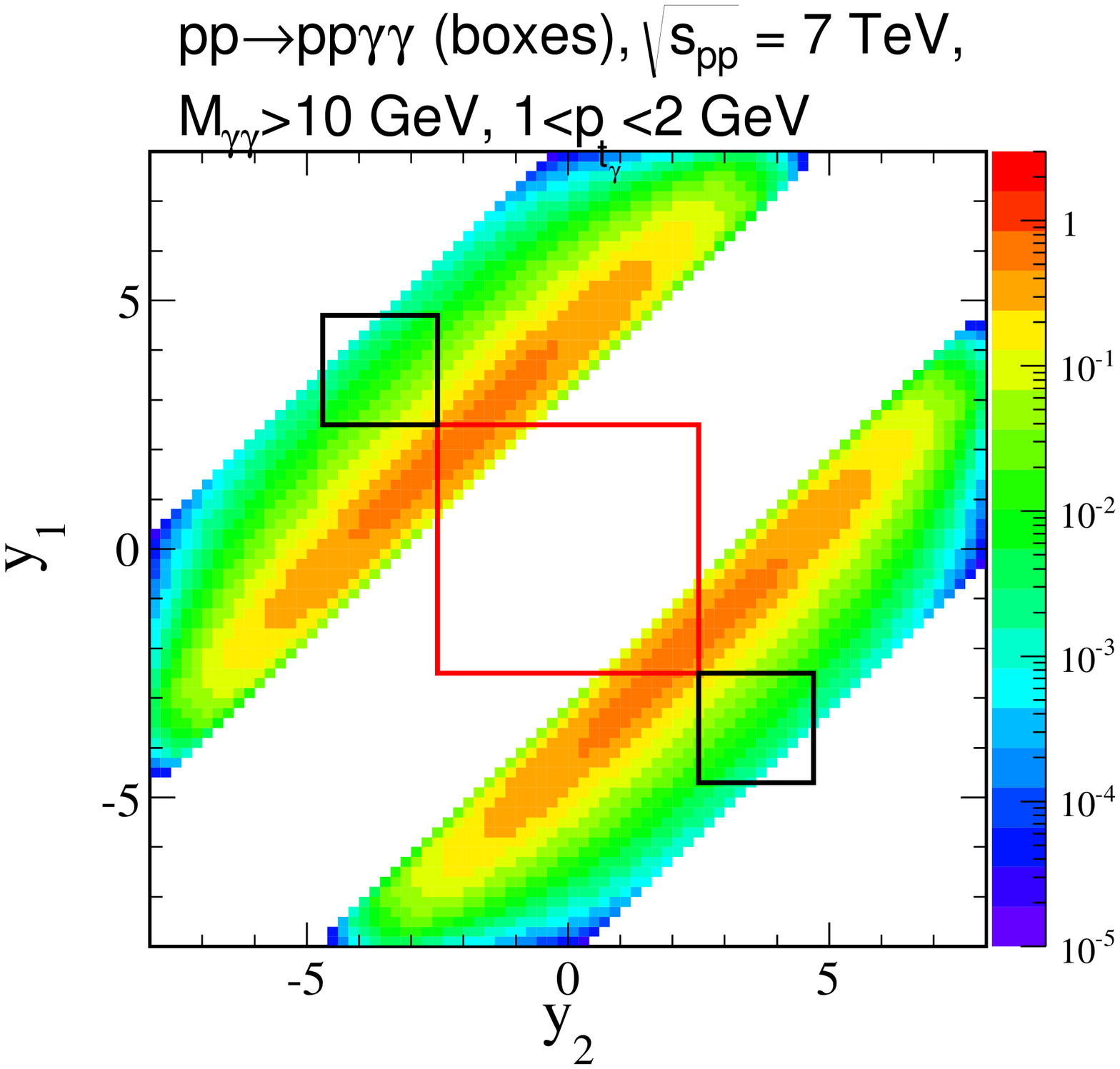}
\includegraphics[width=5cm]{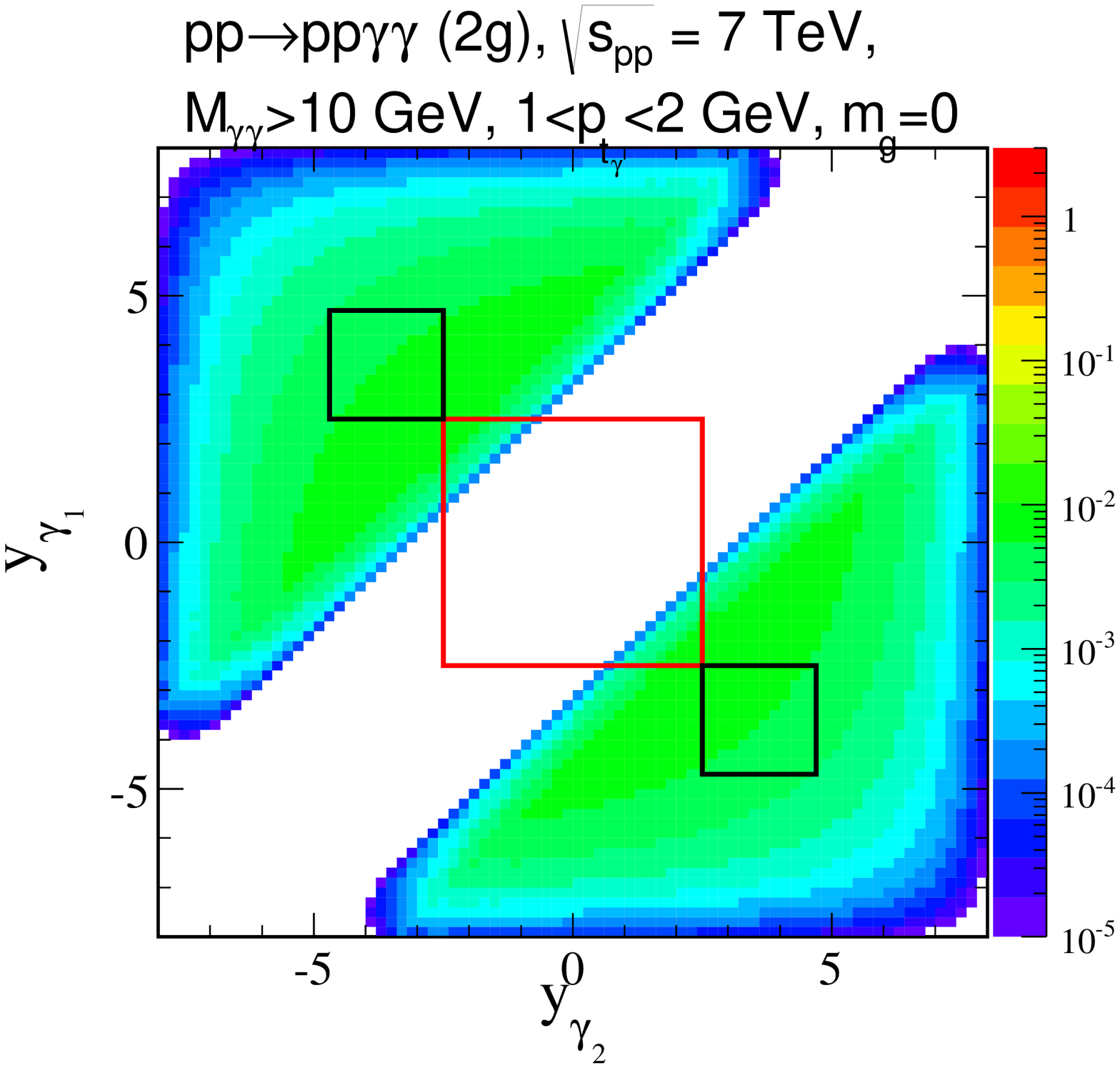}
\includegraphics[width=5cm]{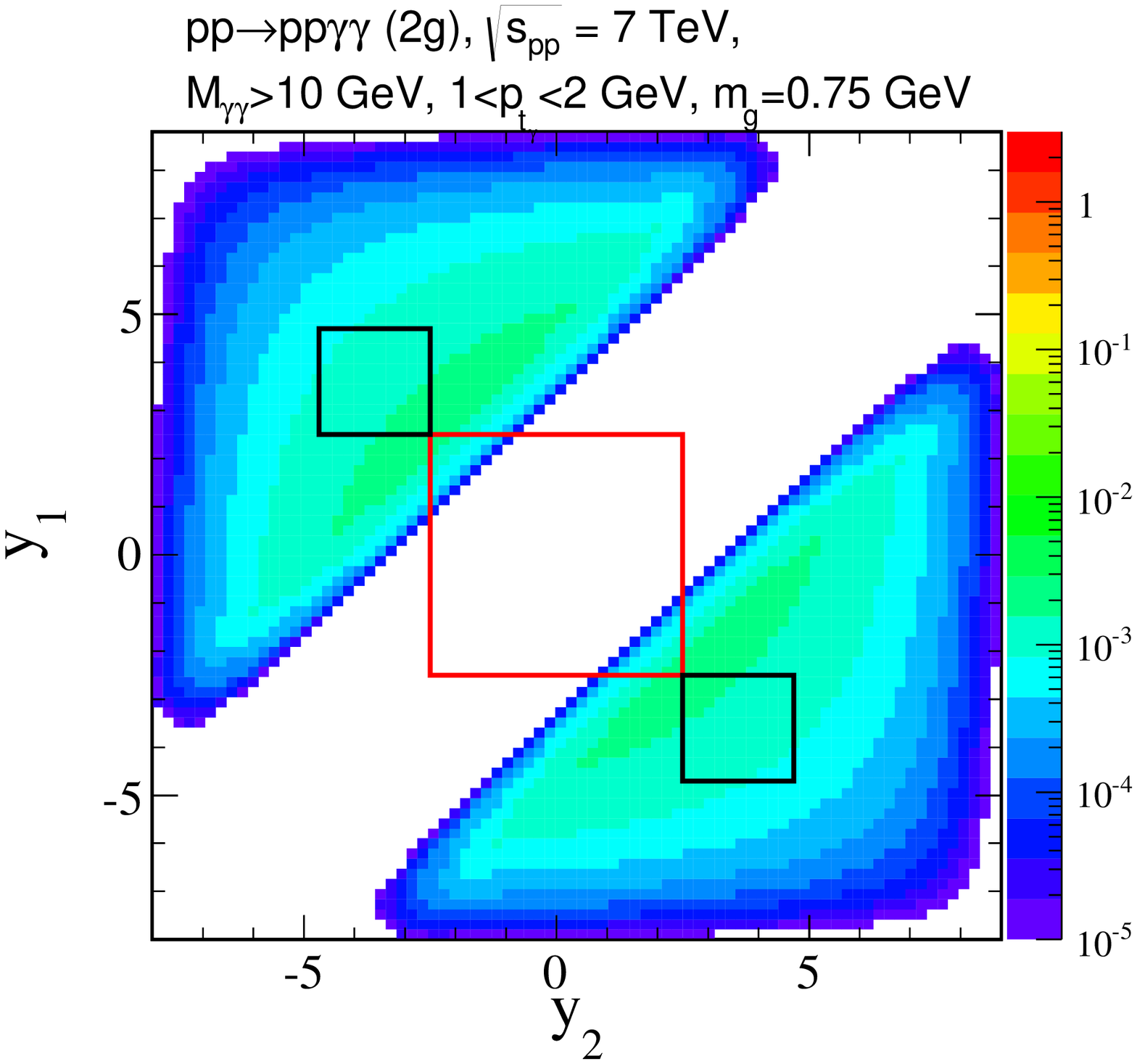}
\caption{Two-dimensional distributions in rapidities of the produced photons
for the box mechanism and the two-gluon echange mechanism
for cuts on $M_{\gamma \gamma}$ and photon transverse momenta specified 
in the figure legend.
We show results both for $m_g =$ 0 and $m_g =$ 0.75 GeV.
}
\label{fig:dsig_dy1dy2}
\end{figure}

\begin{figure}[!h]
\includegraphics[scale=0.35]{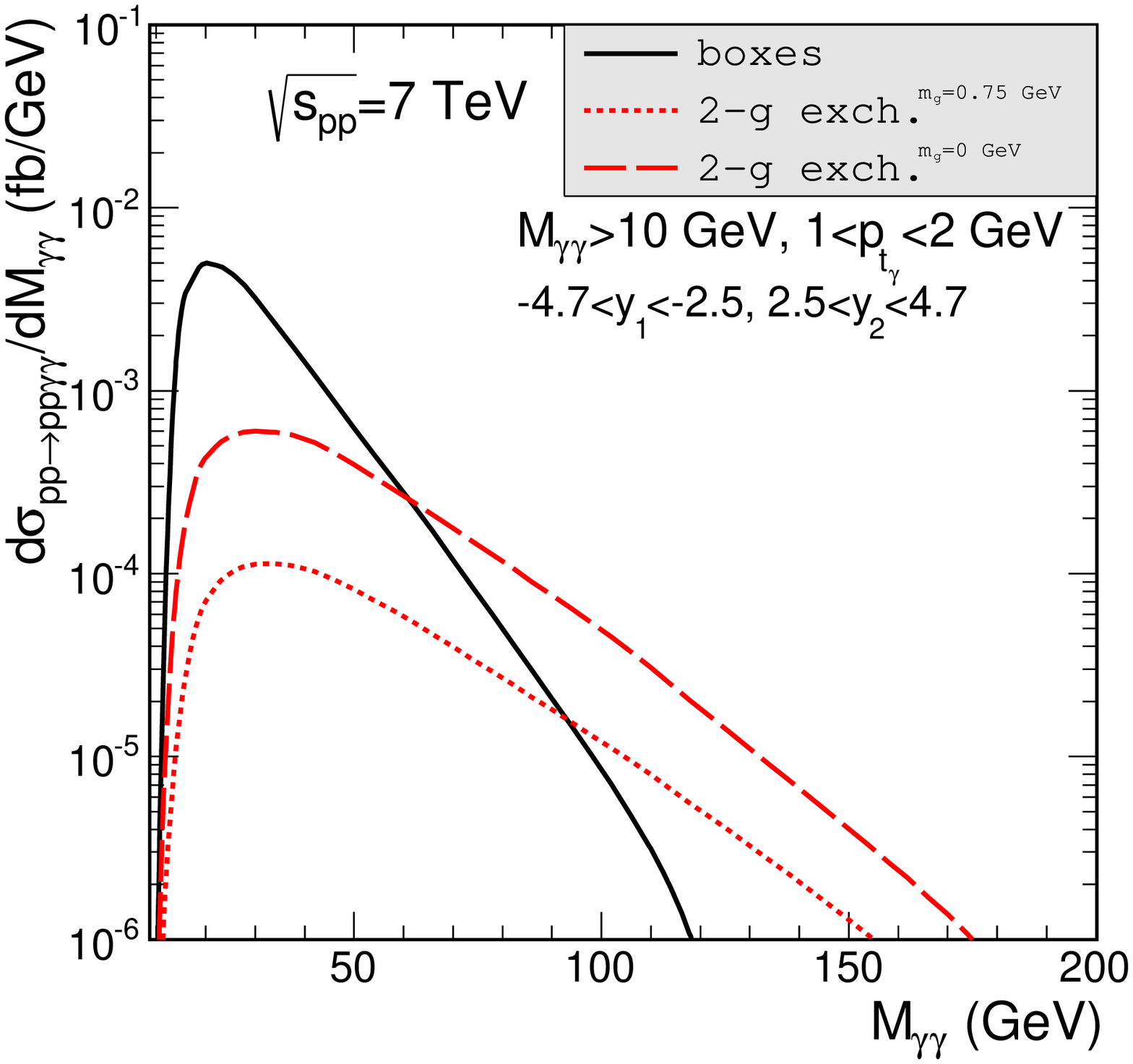}
\includegraphics[scale=0.35]{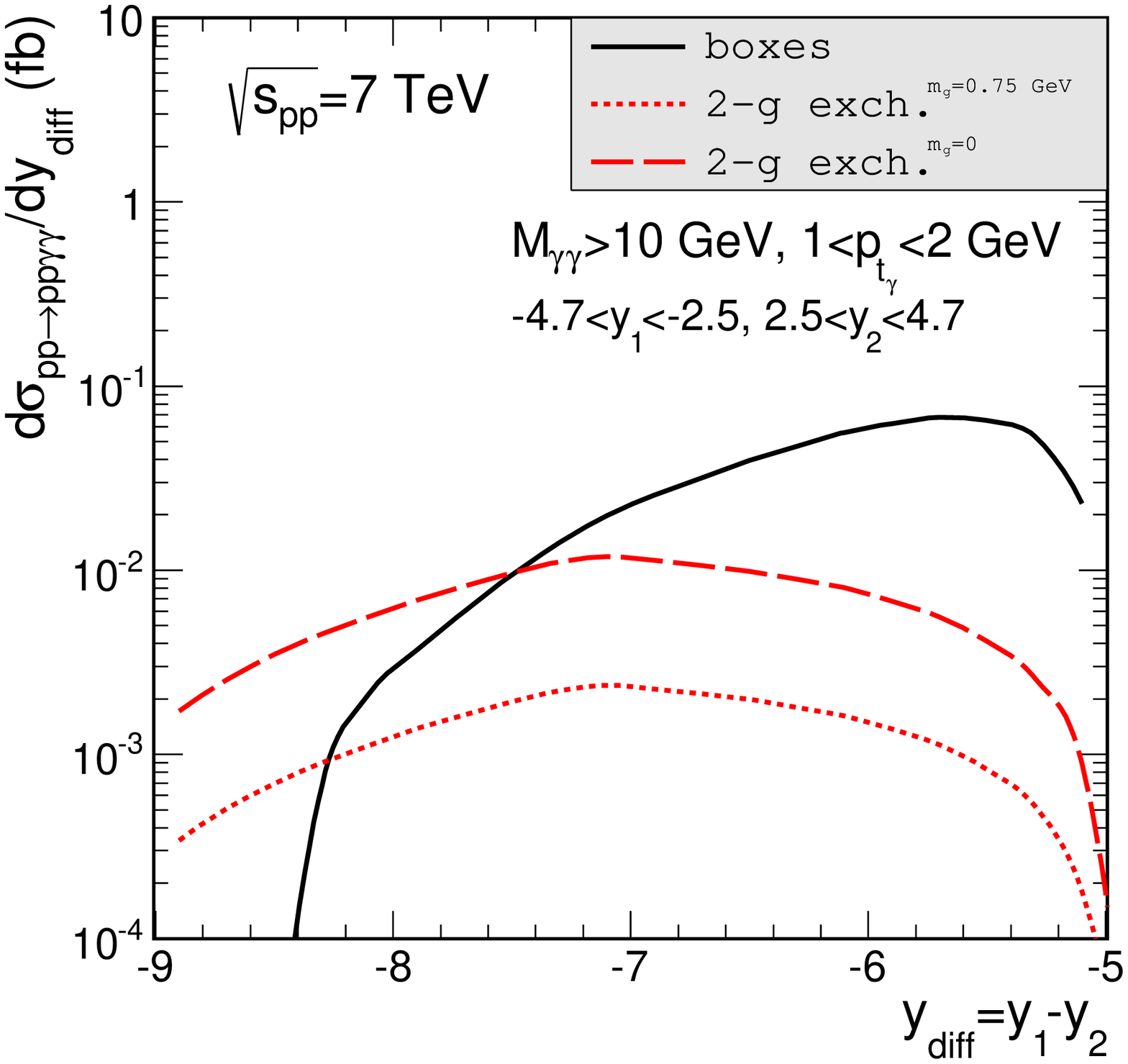}
\caption{Distributions in $M_{\gamma\gamma}$ (left panel) and in 
$y_{diff}$ of the produced photons (right panel)
for $\sqrt{s}$ = 7 TeV (LHC). Here we assumed that one photon 
is measured in one-side calorimeter 
and the second photon in the second-side calorimeter.
Other cuts are specified in the figure legend.
}
\label{fig:dsig_dydiff_calorimeters}
\end{figure}

Let us concentrate for a while on a measurement of photons with 
the help of forward calorimeters (ATLAS or CMS).
To better illustrate the situation in 
Fig.~\ref{fig:dsig_dydiff_calorimeters}
we show distributions in invariant {\texttt{mass of the}} two-photon system and in 
rapidity difference between one photon measured
on one side and the second photon measured on the other side.
The cross section for the $pp \to pp \gamma\gamma$ process  
for the box contribution is larger than that for the two-gluon exchange up to
$M_{\gamma\gamma} = 60$ GeV. For larger values of invariant mass of two photons
the two-gluon exchange contribution (for $m_g=0$) starts to dominate.
For two-gluon exchange naturally the rapidity distances between 
the two photons are large.
The two-gluon exchange contribution becomes larger for rapidity
separations larger than seven or eight units. The corresponding cross
sections are placed in Table 1. The two-gluon exchange contribution
is only a small fraction of fb so the respective measurement would
require large integrated luminosity which may be difficult in the light
of pile-ups which are difficult to handle in the case of exclusive
processes. 
Again the difference between two-gluon exchange contribution
for massive ($m_g=750$ MeV) and massless gluon exchange amounts to almost 
one order of magnitude.

\section{Conclusions}

\begin{table}[h!]
\begin{tabular}{|l|l|r|r|}
\hline
Limitation		& Mechanism		& $\sigma_{PbPb \to Pb Pb \gamma \gamma}$ [nb]	& $\sigma_{pp \to pp \gamma \gamma}$ [fb]
\\ \hline
\hline
$M_{\gamma\gamma}>10$ GeV,	& boxes						&	7.307		&	12.524	\\
1 GeV $<p_{t_\gamma}<$ 2 GeV,		& 2g-exch. ($m_g=0$)			&	1.234		&	0.317	\\
$-8<y_{1}<8$,				& 2g-exch. ($m_g=0.75$ GeV)	&	0.260		& 	0.067	\\	
$-8<y_{2}<8$					& $PbPb \to Pb Pb e^+e^-$ 					&	46 474.000	&				\\	\hline
$M_{\gamma\gamma}>10$ GeV,	& boxes						&	0.063		&	0.105	\\
1 GeV $<p_{t_\gamma}<$ 2 GeV,		& 2g-exch. ($m_g=0$)			&	0.092		&	0.027	\\
$-4.7<y_{1}<-2.5$,			& 2g-exch. ($m_g=0.75$ GeV)	&	0.017		& 	0.005	\\
$2.5<y_{1}<4.7$				& $PbPb \to Pb Pbe^+e^-$ 					&	763.000		&            \\ \hline
\end{tabular}
\caption{ \small
Integrated cross section for the $\gamma\gamma$ production
in lead-lead and proton-proton collisions for LHC energy $\sqrt{s_{NN}}=5.5$ TeV and $\sqrt{s_{pp}}=7$ TeV,
respectively.
We show results for $M_{\gamma\gamma}>10$ GeV and 1 GeV $<p_{t_\gamma}<$ 2 GeV
for full range and for forward calorimeters.
The nuclear cross section is calculated for ultraperipheral collisions of heavy ions.
}
\label{table:cross_sections}
\end{table}

In the present paper we have presented detailed 
formulae for the off-forward two-gluon exchange amplitude(s) for elastic 
photon-photon scattering, including massive quarks and all helicity 
configurations of photons. We have also performed first calculations
of the corresponding component to the elastic photon-photon scattering.
Both distribution in $z = \cos \theta$ and in
transverse momentum of the outgoing photon have been presented.
We have shown that helicity-flip contributions are extremely small
compared to helicity-conserving ones. 
The two-gluon exchange component is rather small at small 
$W_{\gamma \gamma} < 20$ GeV compared to the well known box component.
We have identified a window in photon transverse momentum 
(1 GeV $< p_t <$ 2 GeV) where it may be, however, visible. At higher
$W_{\gamma \gamma}$ energies the region where it wins becomes broader 
(1 GeV $< p_t <$ 5 GeV).
Furthermore the cross section could be enhanced by potential 
BFKL resummation effects. This should be discussed in the future
in more detail.

We have also made predictions for the $A A \to A A \gamma \gamma$
and $p p \to p p \gamma \gamma$ reactions including the previously neglected
two-gluon exchange
component. The calculation for ultraperipheral collisions have been done
in the equivalent photon approximation in the impact parameter space,
while the calculation for proton-proton collisions have been done
as usually in the parton model with elastic photon distributions
expressed in terms of proton electromagnetic form factors.
In both cases we have tried to identify regions of the phase space 
where the two-gluon contribution should be enhanced relatively
to the box contribution. 
The region of large rapidity difference between the two emitted photons
and intermediate transverse momenta
1 GeV $< p_t <$ 2-5 GeV seems optimal in this respect.

However, the resulting cross sections are there rather small and huge
statistics would be required to observe a sign of the two-gluon exchange 
contribution or its BFKL improvement (not yet available).

We have considered also an option to measure
both photons by the forward calorimeters.
It is rather difficult to distinguish photons from electrons in FCALs.
In heavy-ion collisions, in addition, the cross section for 
$AA \to AA e^+ e^-$ is huge, so this option seems not realistic.
In $pp \to pp \gamma \gamma$ case the corresponding background would be smaller
but the signal is also reduced.

\vspace{1.5cm}

{\bf Acknowledments}

We are indebted to Daniel Tapia, Iwona Grabowska-Bo{\l}d and Mateusz Dynda{\l}
for a discussion on possibilities of measuring the here discussed processes
by the CMS and ATLAS Collaborations.
This work was partially supported by the Polish grant 
No. DEC-2014/15/B/ST2/02528 (OPUS)
as well as by the Centre for Innovation and Transfer of Natural Sciences
and Engineering Knowledge in Rzesz\'ow.



\end{document}